\documentstyle[12pt]{article}          % usual 12pt
\newskip\humongous \humongous=0pt plus 1000pt minus 1000pt

\newif\ifdtup

\def\ie{\hbox{\rm i.e.}{}}

\def\abs#1{\left| #1\right|}
\def\pr#1{#1^\prime}

\def\beq{\begin{equation}}
\def\eeq{\end{equation}}
\def\eq{\beq\eeq}
\def\beqn{\begin{eqnarray}}
\def\eeqn{\end{eqnarray}}
\relax

\def\dotx{\dotx{\dot\overline{x}}}

\relax

\catcode`\@=11

\@addtoreset{equation}{section}
\def\theequation{\thesection\arabic{equation}}

\def\@normalsize{\@setsize\normalsize{15pt}\xiipt\@xiipt
\abovedisplayskip 14pt plus3pt minus3pt%
\belowdisplayskip \abovedisplayskip
\abovedisplayshortskip \z@ plus3pt%
\belowdisplayshortskip 7pt plus3.5pt minus0pt}

\def\small{\@setsize\small{13.6pt}\xipt\@xipt
\abovedisplayskip 13pt plus3pt minus3pt%
\belowdisplayskip \abovedisplayskip
\abovedisplayshortskip \z@ plus3pt%
\belowdisplayshortskip 7pt plus3.5pt minus0pt
\def\@listi{\parsep 4.5pt plus 2pt minus 1pt
     \itemsep \parsep
     \topsep 9pt plus 3pt minus 3pt}}

\@twosidetrue
\relax

\catcode`@=12

\catcode`\@=11

\def\section{\@startsection{section}{1}{\z@}{3.5ex plus 1ex minus
   .2ex}{2.3ex plus .2ex}{\large\bf}}

\def\thesection{\arabic{section}.}

\def\appendix{\setcounter{section}{0}
 \def\thesection{APPENDIX \Alph{section}:}
 \def\theequation{\Alph{section}.\arabic{equation}}}
\def\ps@headings{\def\@oddfoot{}\def\@evenfoot{}
\def\@oddhead{\hbox{}\hfill
 \makebox[.5\textwidth]{\raggedright\ignorespaces --\thepage{}--
 \hfill {}}}  %instead of {\rm FERMILAB--Pub--\FERMIPUB}}}
\def\@evenhead{\@oddhead}
\def\subsectionmark##1{\markboth{##1}{}}
}

\ps@headings

\catcode`\@=12

\def\figcap{\section*{Figure Captions\markboth
 {FIGURECAPTIONS}{FIGURECAPTIONS}}\list
 {Fig. \arabic{enumi}:\hfill}{\settowidth\labelwidth{Fig. 999:}
 \leftmargin\labelwidth
 \advance\leftmargin\labelsep\usecounter{enumi}}}
 \relax
\def\tablecap{\section*{Table Captions\markboth
 {TABLECAPTIONS}{TABLECAPTIONS}}\list
 {Table \arabic{enumi}:\hfill}{\settowidth\labelwidth{Table 999:}
 \leftmargin\labelwidth
 \advance\leftmargin\labelsep\usecounter{enumi}}}
 \relax
\def\reflist{\section*{References\markboth
 {REFLIST}{REFLIST}}\list
 {[\arabic{enumi}]\hfill}{\settowidth\labelwidth{[999]}
 \leftmargin\labelwidth
 \advance\leftmargin\labelsep\usecounter{enumi}}}
 \relax

\catcode`\@=11

\def\ps@headings{\def\@oddfoot{}\def\@evenfoot{}
\def\@oddhead{\hbox{}\hfill
 \makebox[.5\textwidth]{\raggedright\ignorespaces --\thepage{}--
 \hfill {}}}    %instead of {\rm FERMILAB--Pub--\FERMIPUB}}}
\def\@evenhead{\@oddhead}
\def\subsectionmark##1{\markboth{##1}{}}
}

\ps@headings

\relax

\relax
\def\pl#1#2#3{{\it Phys. Lett. }{\bf #1}(19#2)#3}
\def\zp#1#2#3{{\it Z. Phys. }{\bf #1}(19#2)#3}
\def\prl#1#2#3{{\it Phys. Rev. Lett. }{\bf #1}(19#2)#3}

\def\pr#1#2#3{{\it Phys. Rev. }{\bf #1}(19#2)#3}
\def\np#1#2#3{{\it Nucl. Phys. }{\bf #1}(19#2)#3}

\relax

\begin{document}
\newcommand\sss{\scriptscriptstyle}
\newcommand\mq{\mbox{$m_{\sss \rm Q}$}}
\newcommand\mug{\mu_\gamma}
\newcommand\mue{\mu_e}
\newcommand\muf{\mu_{\sss F}}
\newcommand\mur{\mu_{\sss R}}
\newcommand\muo{\mu_0}
\renewcommand\pt{\mbox{$p_{\sss \rm T}$}}
\newcommand\as{\alpha_{\sss S}}
\newcommand\ep{\epsilon}
\newcommand\litwo{{\rm Li}_2}
\newcommand\aem{\alpha_{\rm em}}
\newcommand\refq[1]{$^{[#1]}$}
\newcommand\avr[1]{\left\langle #1 \right\rangle}
\newcommand\lambdamsb{
\Lambda_5^{\rm \sss \overline{MS}}
}
\newcommand\qqb{{q\overline{q}}}
\newcommand\asb{\as^{(b)}}
\newcommand\qb{\overline{q}}
\newcommand\sigqq{\sigma_{q\overline{q}}}
\newcommand\fqq{f_{q\qb}}
\newcommand\fqqs{f_{q\qb}^{(s)}}
\newcommand\fqqp{f_{q\qb}^{(c+)}}
\newcommand\fqqm{f_{q\qb}^{(c-)}}
\newcommand\fqqpm{f_{q\qb}^{(c\pm)}}
\newcommand\sigpq{\sigma_{\gamma q}}
\newcommand\mpq{{\cal M}_{\gamma q} }
\newcommand\fpq{f_{\gamma q}}
\newcommand\fpqs{f_{\gamma q}^{(s)}}
\newcommand\fpqp{f_{\gamma q}^{(c+)}}
\newcommand\fpqm{f_{\gamma q}^{(c-)}}
\newcommand\fpqpm{f_{\gamma q}^{(c\pm)}}
\newcommand\fpqtm{{\tilde f}_{\gamma q}^{(c-)}}
\newcommand\sigpg{\sigma_{\gamma g}}
\newcommand\mpg{{\cal M}_{\gamma g} }
\newcommand\fpg{f_{\gamma g}}
\newcommand\fpgs{f_{\gamma g}^{(s)}}
\newcommand\fpgm{f_{\gamma g}^{(c-)}}
\newcommand\fpgtm{{\tilde f}_{\gamma g}^{(c-)}}
\newcommand\siggg{\sigma_{gg}}
\newcommand\mgg{{\cal M}_{gg} }
\newcommand\fgg{f_{gg}}
\newcommand\fggs{f_{gg}^{(s)}}
\newcommand\fggp{f_{gg}^{(c+)}}
\newcommand\fggm{f_{gg}^{(c-)}}
\newcommand\fggpm{f_{gg}^{(c\pm)}}
\newcommand\fggtm{{\tilde f}_{gg}^{(c\pm)}}
\newcommand\fqg{f_{qg}}
\newcommand\epb{\overline{\epsilon}}
\newcommand\thu{\theta_1}
\newcommand\thd{\theta_2}
\newcommand\omxr{\left(\frac{1}{1-x}\right)_{\tilde\rho}}
\newcommand\omyo{\left(\frac{1}{1-y}\right)_{\omega}}
\newcommand\opyo{\left(\frac{1}{1+y}\right)_{\omega}}
\newcommand\lomxr{\left(\frac{\log(1-x)}{1-x}\right)_{\tilde\rho}}
\newcommand\MSB{{\rm \overline{MS}}}
\newcommand\vltm{{\log\frac{-t}{m^2}}}
\newcommand\vlwm{{\log\frac{-u}{m^2}}}
\newcommand\vlpm{{\log\frac{1+\beta}{1-\beta}}}
\newcommand\vlpmq{{\log^2\frac{1+\beta}{1-\beta}}}
\newcommand\ltuno{{\log\frac{-t}{s}}}
\newcommand\ltdue{{\log\frac{-u}{s}}}
\newcommand\softb{{\litwo\frac{2\beta}{1+\beta}
                  -\litwo\frac{-2\beta}{1-\beta}
}}
\def \eq {e_{\sss Q}}
\def \ptg {\mbox{$p_{\sss T}^{Q\overline{Q}}$}}
\def \xf  {\mbox{$x_{\sss F}$}}
\def \dphi{\mbox{$\Delta\phi$}}
\def \dy  {\mbox{$\Delta y$}}
\def \pim {\mbox{$\pi^-$}}
\def \epem {\mbox{$e^+e^-$}}
\def \mc   {\mbox{$m_c$}}
\def \mb   {\mbox{$m_b$}}
\def \mqq   {\mbox{$M_{Q\overline{Q}}$}}
\def \tot   {{\rm tot}}
\def \nn    {\nonumber}
\newcommand\qq{{\sss Q\overline{Q}}}
\newcommand\cm{{\sss CM}}
\input{psfig.sty}
\renewcommand\topfraction{1}       % Max. Fraz. di pagina per float in t
\renewcommand\bottomfraction{1}    % Max. Fraz. di pagina per float in b
\renewcommand\textfraction{0}      % Min. Fraz. di pagina per testo
\setcounter{topnumber}{5}          % Max # float in position t
\setcounter{bottomnumber}{5}       % Max # float in position b
\setcounter{totalnumber}{5}        % Max # float in same page
\setcounter{dbltopnumber}{2}       % Max # large float
\begin{titlepage}
\nopagebreak
\vspace*{-1in}
{\leftskip 11cm
\normalsize
\noindent
\newline
CERN-TH.6921/93 \newline
GEF-TH-15/1993

}
\vfill
\begin{center}
{\large \bf
 Heavy-Quark Correlations in Photon-Hadron Collisions
}
\vfill
{\bf Stefano Frixione}
\vskip .3cm
{Dip. di Fisica, Universit\`a di Genova, and INFN, Sezione di Genova,
Genoa, Italy}\\
\vskip .6cm
{\bf Michelangelo L. Mangano}
\vskip .3cm
{INFN, Scuola Normale Superiore and Dipartimento di Fisica, Pisa, Italy}\\
\vskip .6cm
{\bf Paolo Nason\footnotemark}
\footnotetext{On leave of absence from INFN, Sezione di Milano, Milan, Italy.}
and
{\bf Giovanni Ridolfi\footnotemark}
\footnotetext{On leave of absence from INFN, Sezione di Genova, Genoa, Italy.}
\vskip .3cm
{CERN TH-Division, CH-1211 Geneva 23, Switzerland}
\end{center}
\vfill
\nopagebreak
\begin{abstract}
{\small
We describe a next-to-leading-order calculation of the
fully exclusive parton cross section at next-to-leading order
for the photoproduction of heavy quarks.
We use our result to compute quantities of interest
for current fixed-target experiments.
We discuss heavy-quark total cross sections, distributions, and
correlations.
}
\end{abstract}
\vfill
CERN-TH.6921/93 \newline
June 1993    \hfill
\end{titlepage}
\section{Introduction}
Heavy-quark photoproduction is a phenomenon of considerable interest. It is
closely related to the hadroproduction phenomenon, but it is also considerably
simpler, since the incoming photon is a much better understood object than an
incoming hadron. Aside from being a good testing ground of our understanding of
perturbative QCD, it is also a probe of the structure of the target hadron. In
fact, it has been often pointed out that heavy-quark photoproduction is a
viable way to measure the gluon structure function in the
proton\refq{\ref{HeraWorkshop}}.

Radiative corrections to the single-inclusive photoproduction of heavy quarks
have been first computed in ref.~[\ref{EllisNason}]. The recent work of
ref.~[\ref{SmithNeerven}] has confirmed the first computation, thus making the
photoproduction cross section up to order ${\cal O}(\aem\as^2)$ a
well-established result. From the next-to-leading-order computations the
following facts have emerged. First of all, the photoproduction cross section
receives more moderate next-to-leading corrections than the hadroproduction
case. This result has improved the consistency of the data on charm production
with the theoretical computation. In fact, before the radiative corrections
were known, it was difficult to accommodate the experimentally observed
hadroproduction and photoproduction cross sections with the same value of the
charm quark mass, the first one requiring much smaller masses.

A large amount of experimental information is available on photoproduction of
heavy flavours\refq{\ref{expt}}. Comparison between theory and experiments has
not gone much further than the total cross section. This is a consequence of
the fact that only charm production data have been available, and that the
single-inclusive charm spectrum is strongly modified by non-perturbative
effects. There is reasonable hope that, by looking at more exclusive
distributions, we could learn more from photoproduction results. Modern
fixed-target photoproduction experiments have the capability to study
correlations between the heavy quark and antiquark. Furthermore, at the $ep$
collider HERA, a large charm and bottom cross section is expected. It is clear,
therefore, that in order to make progress in the physics of heavy-quark
production, an exclusive next-to-leading-order calculation of the
photoproduction cross section is needed. This may turn out to be useful both in
charm production at fixed-target experiments and at HERA, and in bottom
production at HERA. Since higher-order corrections are moderate even in the
charm case, it is possible that certain charm distributions may be used for QCD
studies.

In this paper we describe a next-to-leading-order computation of the doubly
differential cross section for the photoproduction of heavy-quark pairs. This
computation follows closely the analogous work of refs.~[\ref{MNR1}] and
[\ref{MNR2}] for hadroproduction of heavy quarks. Our result is implemented in
the form of a ``parton'' event generator, which can be used to compute any
distribution accurate to the next-to-leading order in the strong coupling
constant. The problems arising from soft and collinear divergences are dealt
with by generating appropriate sequences of correlated events, in such a way
that the cancellation of collinear and soft singularities takes place for any
well-defined physical distribution (i.e. distributions that are insensitive to
soft and collinear emission). The advantage of this method (developed for the
first time in ref.~[\ref{ZZ}]) is that it does not require any artificial
regularization of the cross section for producing the quark-antiquark pair plus
a light parton. A detailed description of this method is given in
ref.~[\ref{MNR1}]. In what follows we will describe the photoproduction
calculation, with some emphasis on the differences with the hadroproduction
case.

This paper is organized as follows. In Section~2 we give a general description
of the calculation. Some subtleties arise in the photoproduction calculation,
which have to do with factorization scale choices. We discuss these problems in
Section~3. Some phenomenological applications of our result have already been
given in ref.~[\ref{FMNR1}], where a particular doubly differential cross
section (of interest to the extraction of the gluon density from heavy-quark
photoproduction data) is studied. In this work, we limit ourselves to the study
of fixed-target photoproduction. More detailed studies of heavy quark
production at HERA will be given in future works\refq{\ref{FMNR3}}. In
Section~4 we discuss the total cross section, and in Section~5 we discuss the
differential distributions in fixed-target experiments.
\section{Description of the calculation}
The
partonic subprocesses relevant for heavy-quark photoproduction at
order $\aem\as^2$ are the two-body process
\beq
\gamma\, g \to Q\bar{Q}
\eeq
and the three-body processes
\beqn
\label{realsubprocesses}
\gamma\, g &\to& Q\bar{Q} g \nonumber \\
\gamma\, q &\to& Q\bar{Q} q.
\eeqn
We will describe the two-body process in terms of the quantities
\beqn
s&=&(p_1+p_2)^2  \nonumber \\
t&=&(p_1-k_1)^2-m^2=(p_2-k_2)^2-m^2 \nonumber \\
u&=&(p_1-k_2)^2-m^2=(p_2-k_1)^2-m^2,
\eeqn
where $p_1$ is the photon momentum, $p_2$ is the gluon momentum, and
$k_1,\,k_2$ are the momenta of the heavy quark and antiquark, respectively.
We have
$p_1^2=p_2^2=0$ and $k_1^2=k_2^2=m^2$, where $m$ is the mass
of the heavy quarks, and $s+t+u=0$ (notice that the definition of $t$
and $u$ is not the conventional one).

We will use dimensional regularization to deal with the divergences
appearing in intermediate steps of the calculation. For this reason,
we will need the expressions of phase spaces in $d=4-2\ep$ dimensions.
The two-body phase space is given by
\beq
d\Phi_2=\frac{2^{2\ep}}{\Gamma(1-\ep)}
\left(\frac{4\pi}{s}\right)^\ep
\frac{1}{16\pi}\beta^{1-2\ep}\sin^{-2\ep}\theta_1 d\cos\theta_1,
\label{dphi2}
\eeq
where $\beta=\sqrt{1-\rho}$, $\rho=4m^2/s$ and $\theta_1$ is the angle between
$\vec{p}_1$ and $\vec{k}_1$ in the centre-of-mass system of the incoming
partons. Therefore,
\beq
t=-\frac{s}{2}(1-\beta\cos\theta_1).
\eeq

The three-body processes are characterized by five independent scalar
quantities:
\beqn
s&=&(p_1+p_2)^2 \nonumber \\
t_k&=&(p_1-k)^2 \nonumber \\
u_k&=&(p_2-k)^2 \nonumber \\
q_1&=&(p_1-k_1)^2-m^2 \nonumber \\
q_2&=&(p_2-k_2)^2-m^2
\eeqn
where $p_1$ is the photon momentum, $p_2$ is the momentum of the incoming
parton, $k_1$ and $k_2$
are the momenta of the heavy quark and antiquark, respectively, and
$k$ is the momentum of the emitted light parton.
We will often use the variable $s_2$, the invariant mass of the
heavy quark-antiquark pair, which is related to our independent
invariants through
\beq
s_2 = (k_1+k_2)^2=s+t_k+u_k.
\label{invariants}
\eeq

It will be convenient to introduce variables $x$ and $y$, where
$x=s_2/s$ and $y$ is the cosine of the angle between $\vec{p_1}$
and $\vec{k}$ in the centre-of-mass system of the incoming partons.
We have
\beq
\rho\;\leq\;x\;\leq\;1, \ \ \ \
-1\;\leq\;y\;\leq\;1
\eeq
and
\beq
t_k=-\frac{s}{2}(1-x)(1-y), \;\;\;
u_k=-\frac{s}{2}(1-x)(1+y).
\label{tandu}
\eeq

In the centre-of-mass frame of the $Q{\bar Q}$ system,
our four-momenta are given by
\beqn
p_1&=&p^0_1\;(1,0,0,1) \nonumber \\
p_2&=&p^0_2\;(1,0,\sin\psi,\cos\psi) \nonumber \\
k&=&k^0\;(1,0,\sin\psi',\cos\psi') \nonumber \\
k_1&=&\frac{\sqrt{s_2}}{2}\;
(1,\beta_x\sin\theta_2\sin\theta_1,
\beta_x\cos\theta_2\sin\theta_1,\beta_x\cos\theta_1) \nonumber\\
k_2&=&\frac{\sqrt{s_2}}{2}\;
(1,-\beta_x\sin\theta_2\sin\theta_1,
-\beta_x\cos\theta_2\sin\theta_1,-\beta_x\cos\theta_1),
\eeqn
where
\beqn
p_1^0=\frac{s+t_k}{2\sqrt{s_2}},\;\;
p_2^0&=&\frac{s+u_k}{2\sqrt{s_2}},\;\;
k^0=-\frac{t_k+u_k}{2\sqrt{s_2}}\nonumber \\
\cos\psi&=&1-\frac{s}{2p_1^0p_2^0},\;\;\;\sin\psi>0
\nonumber \\
\cos\psi'&=&1+\frac{t_k}{2p_1^0k^0},\;\;\;\sin\psi'>0
\nonumber \\
\beta_x&=&\sqrt{1-\frac{4m^2}{sx}}.
\label{psidef}
\eeqn
The two remaining independent
invariants $q_1,q_2$ are given by
\beqn
q_1&=&-\frac{s+t_k}{2}(1-\beta_x\cos\theta_1) \nonumber \\
q_2&=&-\frac{s+u_k}{2}
(1+\beta_x\cos\theta_2\sin\theta_1\sin\psi
  +\beta_x\cos\theta_1\cos\psi).
\label{q1q2}
\eeqn
Now all invariants are expressed in terms of $x,\,y,\,\theta_1,\,
\theta_2$ and $s$ through eqs.~(\ref{tandu}), (\ref{psidef})
and (\ref{q1q2}).

The three-body phase space in terms of the variables $x,y,\theta_1,\theta_2$
is given by
\beq
d\Phi_3=H N d\Phi_2^{(x)}
\frac{s^{1-\ep}}{2\pi}
(1-x)^{1-2\ep}(1-y^2)^{-\ep}dy
\sin^{-2\ep}\theta_2 d\theta_2,
\eeq
where
\beqn
H&=&\frac{\Gamma(1-\ep)}{\Gamma(1+\ep)\Gamma(1-2\ep)}
=1-\frac{\pi^2}{3}\ep^2+{\cal O}(\ep^3) \\
N&=&\frac{(4\pi)^{\ep}}{(4\pi)^2}\Gamma(1+\ep)\label{N}
\eeqn
and
\beq
d\Phi_2^{(x)}=
\frac{2^{2\ep}}{\Gamma(1-\ep)}
\left(\frac{4\pi}{sx}\right)^\ep
\frac{1}{16\pi}\beta_x^{1-2\ep}\sin^{-2\ep}\theta_1 d\cos\theta_1 dx.
\eeq
Both $\theta_1$ and $\theta_2$ range between $0$ and $\pi$.

We are now ready to compute the cross section for
the real emission processes of eq.~(\ref{realsubprocesses}).
The technique is the same as that used in ref.~[\ref{MNR1}].
We begin with the subprocess $\gamma g \to Q{\bar Q}g$. The cross section
(in $d$ space-time dimensions) is given by
\beqn
d\sigpg^{(r)} &=& \mpg^{(r)}(s,t_k,u_k,q_1,q_2) d\Phi_3 \\
\mpg^{(r)}(s,t_k,u_k,q_1,q_2)&=& \frac{1}{2s}\frac{1}
{\left[2(1-\ep)\right]^2(N_{\sss C}^2-1)}
\sum_{\rm spin,color} \abs{{\cal A}_{\gamma g}^{(r)}}^2,
\eeqn
where ${\cal A}_{\gamma g}^{(r)}$ is the invariant amplitude.
The invariant cross section $\mpg^{(r)}$ has singularities in $t_k=0$ and
$u_k=0$, corresponding to soft ($x=1$) and collinear ($y=-1$) gluon emission.
No collinear emission from the photon line takes place
at this order for the $\gamma g\to Q \bar{Q} g$ subprocess, and
therefore $\mpg^{(r)}$ is regular at $y=1$.
It can be shown that the leading soft singularity behaves like $1/(1-x)^2$, and
that no double poles appear in $t_k$ and $u_k$.
Therefore the function
\beq
\label{fpgdef}
\fpg(x,y,\thu,\thd) =
4 t_k u_k \mpg^{(r)}(s,t_k,u_k,q_1,q_2)
\eeq
is regular for $y=-1$ and $x=1$ (the dependence of $\fpg$ upon $s/m^2$
is not explicitly shown).
Using eqs.~(\ref{tandu}) we get
\beq
\mpg^{(r)}(s,t_k,u_k,q_1,q_2) =
\frac{\fpg(x,y,\thu,\thd)}{s^2(1-x)^2(1-y^2)}.
\eeq
The three-body contribution to our cross section, including
the phase space, is then given by
\beq
d\sigpg^{(r)}=
HNd\Phi_2^{(x)}\frac{s^{-1-\ep}}{2\pi}
dy\sin^{-2\ep}\thd d\thd
(1-x)^{-1-2\ep}(1-y^2)^{-1-\ep}\fpg(x,y,\thu,\thd).
\label{threebody}
\eeq
We can now use the following expansions, valid for small $\ep$
\beqn
(1-x)^{-1-2\ep}&=&-\frac{\tilde\beta^{-4\ep}}{2\ep}\delta(1-x)
+\omxr - 2\ep \lomxr \nonumber \\
&&+{\cal O}(\ep^2) \\
(1-y^2)^{-1-\ep}&=&-[\delta(1+y)+\delta(1-y)]\frac{(2\omega)^{-\ep}}{2\ep}
\nonumber \\
&&+\frac{1}{2}\left[\omyo+\opyo\right]+{\cal O}(\ep),
\label{expansion}
\eeqn
where the distributions in round brackets
are defined according to the prescriptions
\beqn
\int_{\tilde\rho}^1 h(x)\omxr dx &=&
\int_{\tilde\rho}^1 \frac{h(x)-h(1)}{1-x} dx
\nonumber\\
\qquad \int_{\tilde\rho}^1 h(x)\lomxr dx &=&
\int_{\tilde\rho}^1 \left[h(x)-h(1)\right] \frac{\log(1-x)}{1-x} dx
\nonumber\\
\int_{1-\omega}^{1} h(y)\omyo dy &=&
\int_{1-\omega}^1 \frac{h(y)-h(1)}{1- y} dy
\nonumber\\
\int_{-1}^{-1+\omega} h(y)\opyo dy &=&
\int_{-1}^{-1+\omega} \frac{h(y)-h(-1)}{1+y} dy ,
\label{prescriptions}
\eeqn
for any test function $h(x)$.
We define $\tilde\beta=\sqrt{1-\tilde\rho}$. The parameters $\tilde\rho$
and $\omega$ should be chosen within the ranges
\beq
\rho \leq \tilde\rho < 1 \; , \quad 0 < \omega \leq 2 .
\eeq
The final results will not depend on the particular values
chosen for $\tilde\rho$ and $\omega$,
but different choices can lead to better convergence in the numerical
programs, as discussed in ref.~[\ref{MNR1}]. We obtain
\beqn
d\sigpg^{(r)}&=&d\sigpg^{(s)}+
HNd\Phi_2^{(x)}\frac{s^{-1-\ep}}{2\pi}
dy\sin^{-2\ep}\thd d\thd
\nonumber \\ &&\times
\left[\omxr-2\ep\lomxr\right]
(1-y^2)^{-1-\ep}\fpg(x,y,\thu,\thd),
\label{xpgdecomp}
\eeqn
where
\beqn
d\sigpg^{(s)}&=&
HNd\Phi_2^{(x)}\frac{s^{-1-\ep}}{2\pi}
dy\sin^{-2\ep}\thd d\thd
\nonumber \\ &&\times
\left[-\frac{\tilde\beta^{-4\ep}}{2\ep} \delta(1-x) \right]
(1-y^2)^{-1-\ep}\fpg(x,y,\thu,\thd).
\label{sigmas}
\eeqn
The details of the calculation of the soft component of the cross section,
$d\sigpg^{(s)}$, are given in Appendix~A.
Equation (\ref{sigmas}) can be explicitly integrated over $x$, $y$ and
$\thd$ to obtain
\beq
d\sigpg^{(s)}=
-HN d\Phi_2\frac{1}{4\pi\ep}
s^{-1-\ep}\tilde\beta^{-4\ep}\fpgs(\thu),
\label{sigpgs}
\eeq
where the function $\fpgs(\thu)$ is given in Appendix~A.

We now expand $(1-y^2)^{-1-\ep}$ in the second
term of eq.~(\ref{xpgdecomp}), observing that we only
need the expansion up to order $\ep^0$.
As noticed above, $\mpg^{(r)}$ is regular at $y=1$, and therefore
the term proportional to $\delta(1-y)$ gives no contribution.
We get
\beq
d\sigpg^{(r)}=d\sigpg^{(s)}+d\sigpg^{(c-)}+d\sigpg^{(f)},
\eeq
where
\beqn
d\sigpg^{(c-)}&=&
Nd\Phi_2^{(x)}\frac{s^{-1-\ep}}{2\pi}
dy\sin^{-2\ep}\thd d\thd
\left[\omxr-2\ep\lomxr\right]
\nonumber \\ &&\times
\left[-\frac{(2\omega)^{-\ep}}{2\ep}\delta(1+y)\right]
\fpg(x,y,\thu,\thd)
\label{sigmacm}
\eeqn
and
\beqn
d\sigpg^{(f)}&=&N\frac{s^{-1}}{64\pi^2}\beta_x d\cos\thu d\thd dy dx
\nonumber \\ &&\times
\omxr\left[\omyo+\opyo\right]\fpg(x,y,\thu,\thd).
\label{sigmaf}
\eeqn
The technique for the evaluation of the collinear limit of the invariant
cross section is described in detail in
ref.~[\ref{MNR1}]. The result in our case is
\beq
\label{sigmacm1}
d\sigpg^{(c-)}=-N\frac{s^{-1-\ep}}{4\ep}\left(\frac{2}{\omega}\right)^\ep
d\Phi_2^{(x)}
\left[\omxr-2\ep\lomxr\right] \fpgm(x,\thu),
\eeq
where
\beq
\label{fcm}
\fpgm(x,\thu)= 64\pi C_{\sss A}\asb s \left(1-x\right)
\left[\frac{x}{1-x}+\frac{1-x}{x}+x(1-x)\right]\mpg^{(b)}(xs,q_1).
\eeq
Here $\asb=\as\mu^{2\ep}$ is the dimensionful coupling constant in $d$
dimensions (the suffix $(b)$ stands for bare), and $\mpg^{(b)}(s,t)$
is the invariant cross section for $\gamma g\to Q {\bar Q}$ at the Born
level,
\beq
\label{sigmabornpg}
d\sigpg^{(b)} = \mpg^{(b)}(s,t)d\Phi_2.
\eeq
The explicit expression for $\mpg^{(b)}$ is given in Appendix A.
The term in the square bracket in eq.~(\ref{fcm}) is, up to a
factor $2C_{\sss A}$, the gluon-gluon Altarelli-Parisi splitting
function in $d$ dimensions for $x<1$,
and the Born cross section is taken in $d$ dimensions.

With the usual definition
\beq
\frac{1}{\epb}=\frac{1}{\ep}-\gamma_E+\log(4\pi),
\eeq
we can rewrite eq.~(\ref{sigmacm1}) as
\beqn
d\sigpg^{(c-)}&=&-\frac{s^{-\ep}}{\overline\ep}
\left(\frac{2}{\omega}\right)^\ep\frac{C_{\sss A}\asb}{\pi}
\left[\omxr-2\ep\lomxr\right]
\nonumber \\
&&\times
\left[x+\frac{(1-x)^2}{x}+x(1-x)^2\right]
\mpg^{(b)}(xs,q_1) d\Phi_2^{(x)}.
\eeqn
We see that the $1/\ep$ divergence
in the collinear term assumes the form dictated by the factorization
theorem.
According to this factorization theorem\refq{\ref{Factorization}},
any partonic cross section can be written as
\beq
d\sigma_{ij}(p_1,p_2)=\sum_{kl} \int d\hat\sigma_{kl}
(x_1 p_1,x_2 p_2) \Gamma_{ki}(x_1) \Gamma_{lj}(x_2)
dx_1 dx_2,
\label{factform}
\eeq
where
\beq
\label{gammadiv}
\Gamma_{ij}(x)=\delta_{ij}\delta(1-x)-\frac{1}{\overline{\ep}}
\frac{\as}{2\pi} P_{ij}(x)+
\frac{\as}{2\pi} K_{ij}(x)+{\cal O}(\as^2)
\eeq
and $d\hat\sigma$ is free of singularities as $\ep$ goes
to zero. The collinear factors $\Gamma_{ij}(x)$ are usually reabsorbed
into the hadronic structure functions, and only the quantities
$d\hat\sigma_{kl}$ will enter the physical cross
section. The functions $P_{ij}(x)$ are the Altarelli-Parisi kernels.
The functions $K_{ij}(x)$ in eq.~(\ref{gammadiv}) are completely arbitrary,
different choices corresponding to different subtraction schemes.
The choice $K_{ij}(x)=0$, to which we stick in the following,
corresponds to the $\MSB$ subtraction scheme\refq{\ref{BBDM}}.

Expanding eq.~(\ref{factform}) order by order in perturbation theory, we find
for our case
\beq
d\hat\sigpg(p_1,p_2)= d\sigpg(p_1,p_2) +
\frac{1}{\overline{\ep}}\frac{\as}{2\pi}P_{gg}(x)
\mpg^{(b)}(xs,q_1) d\Phi_2^{(x)},
\label{sigmahat}
\eeq
where
\beqn
P_{gg}(x)&=&2C_{\sss A} \left[\frac{x}{(1-x)}_+
+\frac{1-x}{x} + x(1-x)\right] + 2\pi b_0 \delta(1-x)
\nonumber \\
&=&2C_{\sss A} \left[\frac{x}{(1-x)}_{\tilde\rho} +
\frac{1-x}{x} + x(1-x)\right]
+ (2\pi b_0+4 C_{\sss A}\log{\tilde\beta}) \delta(1-x),
\nonumber \\
\eeqn
and
\beq
b_0=\frac{11C_{\sss A}-4T_{\sss F} n_{\rm lf}}{12\pi}.
\eeq
Here $n_{\rm lf}$ is the number of light flavours, and for $N_{\sss C}=3$
we have
\beq
C_{\sss A}=3,\quad T_{\sss F}=\frac{1}{2}.
\eeq
The final expression for the short-distance
cross section, after subtraction of the collinear divergences,
eq.~(\ref{sigmahat}), becomes
\beq
d\hat\sigpg = d\sigpg^{(b)} +
 d\hat\sigpg^{(c-)} + d\hat\sigpg^{(s)} + d\sigpg^{(v)}
+d\sigpg^{(f)},
\label{sigpg}
\eeq
where $d\sigpg^{(b)}$ and $d\sigpg^{(f)}$ are as given in
eqs.~(\ref{sigmabornpg}) and (\ref{sigmaf}), respectively,
\beqn
d\hat\sigpg^{(c-)}&=&\phantom{\times}\frac{C_{\sss A}\as}{\pi}
            \left[\left(\log\frac{s}{\muf^2}+\log\frac{\omega}{2}\right)
            \omxr+2\lomxr\right]
\nonumber \\&&
            \phantom{\frac{C_{\sss A}\as}{\pi}}\times
            \left[x+\frac{(1-x)^2}{x} + x(1-x)^2\right]
            \mpg^{(b)}(xs,q_1) d\Phi_2^{(x)}
\label{colpgm}
\\
d\hat\sigpg^{(s)}&=&d\sigpg^{(s)}+
                  \frac{C_{\sss A}\as}{\pi}\frac{1}{\bar\ep}
                  (2\pi b_0 + 4C_{\sss A}\log{\tilde\beta})
                  \mpg^{(b)}(s,t) d\Phi_2.
\label{softerm}
\eeqn
The scale $\mu$, appearing explicitly in the expression of $\asb$, has been set
equal to a scale $\muf$ characteristic of the subtraction
of the singularity due to collinear emission from the
incoming gluon, while $\as$ is taken everywhere at the renormalization
scale $\mur$
(the problem of scale definitions
and choices will be discussed extensively in the next section).
The remaining singularities in $d\hat\sigpg^{(s)}$ are cancelled
by the singularities
in the virtual contribution to the cross section $d\sigpg^{(v)}$.
Therefore the quantity
\beq
d\sigpg^{(sv)} = d\hat\sigpg^{(s)}+d\sigpg^{(v)}
\eeq
is finite as $\ep \to 0$, and so is the full expression for $d\hat\sigpg$.

We now turn to the other three-body subprocess present at the
$\aem\as^2$ level, namely $\gamma q \to Q{\bar Q}q$.
In this case, collinear emission takes place both from the photon
and from the incoming light quark. On the other hand, there is
no order $\aem\as$ contribution to heavy-quark pair production
via $\gamma q$ fusion. Therefore, no
soft singularity is expected. We explicitly checked that this is indeed
the case. The three-body cross section is given by
\beqn
d\sigpq^{(r)} &=& \mpq^{(r)}(s,t_k,u_k,q_1,q_2) d\Phi_3 \\
\mpq^{(r)}(s,t_k,u_k,q_1,q_2)&=& \frac{1}{2s}\frac{1}
{2(1-\ep)N_{\sss C}}
\sum_{\rm spin, color} \abs{{\cal A}_{\gamma q}^{(r)}}^2,
\eeqn
where ${\cal A}_{\gamma q}^{(r)}$ is the invariant amplitude. The invariant
cross section $\mpq^{(r)}$ has singularities for $t_k=0$, corresponding to
collinear ($y=1$) light-quark emission from the photon, or $u_k=0$,
corresponding to collinear ($y=-1$) light-quark emission from the incoming
quark. The function
\beq
\label{fpqdef}
\fpq(x,y,\thu,\thd) =
4t_k u_k
\mpq^{(r)}(s,t_k,u_k,q_1,q_2)
\eeq
is therefore regular for $y=\pm 1$, and vanishes as $(1-x)^2$ for $x\to 1$.
For this reason, the three-body cross section can be rewritten as
\beqn
d\sigpq^{(r)}&=&HNd\Phi_2^{(x)}\frac{s^{-1-\ep}}{2\pi}
dy\sin^{-2\ep}\thd d\thd
\nonumber \\ &&\times
\left[\omxr-2\ep\lomxr\right]
(1-y^2)^{-1-\ep}\fpq(x,y,\thu,\thd).
\label{xpqdecomp}
\eeqn
Expanding $(1-y^2)^{-1-\ep}$ in eq.~(\ref{xpqdecomp}) we get
\beq
d\sigpq^{(r)}=d\sigpq^{(c+)}+d\sigpq^{(c-)}+d\sigpq^{(f)},
\eeq
where
\beqn
d\sigpq^{(c\pm)}&=&
Nd\Phi_2^{(x)}\frac{s^{-1-\ep}}{2\pi}
dy\sin^{-2\ep}\thd d\thd
\left[\omxr-2\ep\lomxr\right]
\nonumber \\ &&\times
\left[-\frac{(2\omega)^{-\ep}}{2\ep}\delta(1\mp y)\right]
\fpq(x,y,\thu,\thd)
\label{sigmacpm}
\eeqn
and
\beqn
d\sigpq^{(f)}&=&N\frac{s^{-1}}{64\pi^2}\beta_x d\cos\thu d\thd dy dx
\nonumber \\ &&\times
\omxr\left[\omyo+\opyo\right]\fpq(x,y,\thu,\thd).
\label{sigmapqf}
\eeqn
Performing the $y$ and $\thd$ integrations in eq.~(\ref{sigmacpm}) we
obtain
\beq
\label{sigmacpm1}
d\sigpq^{(c\pm)}=-N\frac{s^{-1-\ep}}{4\ep}\left(\frac{2}{\omega}\right)^\ep
d\Phi_2^{(x)}
\left[\omxr-2\ep\lomxr\right] \fpqpm(x,\thu).
\eeq
The explicit form of $\fpqpm(x,\thu)$ can be obtained following
ref.~[\ref{MNR1}]. The results are
\beq
\fpqp(x,\thu)=
32 \pi\aem  e_q^2 s (1-x){\rm P}_{q\gamma}(x)
{\cal M}^{(b)}_{q\bar{q}}(xs,q_2)\, ,
\eeq
where $e_q$ is the charge of the emitted light quark in electron
charge units, ${\cal M}^{(b)}_{q\bar{q}}(s,t)$ is
the lowest-order invariant cross section for $q\bar{q}\to Q\bar{Q}$,
\beq
d\sigqq^{(b)}={\cal M}_{q\qb}^{(b)}(s,t) d\Phi_2,
\eeq
and
\beq
{\rm P}_{q\gamma}(x)\,=\,
N_{\sss C}\left[x^2+(1-x)^2-2x(1-x)\ep\right]
\eeq
is the Altarelli-Parisi splitting function in $4-2\ep$ dimensions
entering the probability of finding a quark in a photon,
which is clearly equal to the splitting
function of a quark in a gluon, up to a colour factor $T_{\sss F}/N_{\sss C}$,
due to the $\lambda$ matrix in the $gq\bar{q}$ vertex.
For $\fpqm$ we obtain
\beq
\fpqm(x,\thu)=32\pi \as s (1-x) {\rm P}_{gq}(x)\mpg^{(b)}(xs,q_1),
\eeq
where
\beq
{\rm P}_{gq}(x)=C_{\sss F} \frac{1+(1-x)^2-\ep x^2}{x}
\eeq
is the Altarelli-Parisi splitting function of a gluon into a quark
in $4-2\ep$ dimensions.
The subtraction of collinear singularities takes place as discussed for the
previous case. We just give our final formulae,
\beq
d\hat\sigpq = d\hat\sigpq^{(c+)} +
 d\hat\sigpq^{(c-)} +d\sigpq^{(f)},
\label{sigpq}
\eeq
where $d\sigpq^{(f)}$ is given in eq.~(\ref{sigmapqf}),
\beqn
d\hat\sigpq^{(c+)}&=&\frac{N_{\sss C}\aem e_q^2}{2\pi}
{\cal M}_{q\qb}^{(b)}(xs,q_2) d\Phi_2^{(x)}
\nonumber\\
&&\times
\Bigg[2x(1-x)+\left(x^2+(1-x)^2\right)
\left(\log\frac{s}{\mug^2}+\log\frac{\omega}{2}+2\log(1-x)\right)
\Bigg]
\label{colpqp}
\nonumber\\
\\
d\hat\sigpq^{(c-)}&=&\frac{C_{\sss F}\as}{2\pi}
\mpg^{(b)}(xs,q_1) d\Phi_2^{(x)}\nonumber \\
&&\times
\Bigg[x+\frac{1+(1-x)^2}{x}
\left(\log\frac{s}{\muf^2}+\log\frac{\omega}{2}+2\log(1-x)\right)
\Bigg].
\label{colpqm}
\eeqn
Notice that collinear emission from the photon is characterized by a
scale $\mug$ which is {\it a priori} different from the hadronic
factorization scale $\muf$.
The quantities
$d\sigpg^{(f)}$ in eq.~(\ref{sigpg})
and $d\sigpq^{(f)}$ in eq.~(\ref{sigpq}) can be found in the
literature\refq{\ref{EllisKunszt}}, and we did not need to explicitly
evaluate them. The quantity $d\sigpg^{(v)}$ in eq.~(\ref{sigpg})
was obtained from the authors of ref.~[\ref{EllisNason}].

The analytical results presented here are implemented as a parton
event generator, written in FORTRAN.
The interested reader can obtain the code from the authors.

\section{Scales in the photoproduction process}
Heavy-quark photoproduction differs from hadroproduction
in the treatment of collinear singularities.
In fact, when a light parton is collinearly emitted by
the incoming photon, the subtracted term is a signal from the
non-perturbative region where the photon splits into quarks and
gluons before interacting with the partons in the hadronic
target. This fact is taken into account by inserting in the
photon-hadron cross section a contribution in which the photon
is formally treated as a hadron (the so-called hadronic or resolved
photon component, to distinguish it from the point-like or pure-photon
component, in which the photon directly couples with the
partons of the hadronic target).
The photon structure functions will also depend upon the momentum scale
$\mu_\gamma$ at which the collinear singularities of the photon leg
are subtracted.
Neither the point-like nor the hadronic components are separately
independent of $\mu_\gamma$, because the subtracted term
in the point-like component is responsible for the redefinition
of the photon structure functions in the hadronic component.

Let us consider first the heavy-quark production process
of an on-shell photon colliding with a hadron $H$ at centre-of-mass
energy $\sqrt{S}$.
In order to clarify the r\^ole of the various scale dependences in the
process, we write the ${\cal O}(\aem\as^2)$ cross section in the following form
\beqn
\sigma^{(\gamma {\sss H})}_{\sss {Q\bar{Q}}}(S)&=&
\sum_i\int {dx}\,f^{(H)}_i(x,\muf)\hat{\sigma}_{\gamma i}
  (xS,\as(\mur),\mur,\muf,\mug) \nonumber \\
&&+ \sum_{ij}\int dx_1\,dx_2\, f_i^{(\gamma)}(x_1,\mug,\muf^\prime)
f^{(H)}_j(x_2,\muf^\prime) \hat{\sigma}_{ij}
  (x_1 x_2 S,\as(\mur^\prime),\mur^\prime,\muf^\prime)\nonumber \\
  &&+{\cal O}(\aem\as^3) \label{sigmascales}
\eeqn
with
\beqn
\hat{\sigma}_{\gamma i}
  (s,\as(\mur),\mur,\muf,\mug)&=&
   \aem\as(\mur)\hat{\sigma}^{(0)}_{\gamma i}
  (s)+\aem\as^2(\mur)\hat{\sigma}^{(1)}_{\gamma i}
  (s,\mur,\muf,\mug) \nonumber \\
\hat{\sigma}_{ij}
  (s,\as(\mur),\mur,\muf)&=&
   \as^2(\mur)\hat{\sigma}^{(0)}_{ij}
  (s)+\as^3(\mur)\hat{\sigma}^{(1)}_{ij}
  (s,\mur,\muf,\mug).
\eeqn
Here $\mur$ and $\mur^\prime$ are renormalization scales,
$\muf$ and $\muf^\prime$ are factorization scales for collinear
singularities arising from strong interactions,
and $\mug$ is a
factorization scale for collinear singularities arising from
the electromagnetic vertex.
If one wanted to extend eq.~(\ref{sigmascales}) to even higher
orders, one should also include an explicit dependence of the structure
functions upon the renormalization scale.
At the order we are considering,
the renormalization scale in the structure functions can be kept equal
to the factorization scale, as is usually done.
The left-hand side is independent of all the scales up to terms of
order $\aem\as^3$, provided the
parton density functions obey the appropriate evolution equations.
The hadronic and photonic parton densities obey the usual
Altarelli-Parisi equations in $\muf$. In addition, the photonic
parton densities have also an inhomogeneous evolution in $\mug$, which,
at the leading order, is given by
\beq \label{inhomogeneous}
\frac{\partial f^{(\gamma)}_i(x,\muf,\mug)}{\partial\log\mug^2}=
\frac{\aem}{2\pi} e_i^2
\left[x^2+(1-x)^2\right] + {\cal O}(\as),
\eeq
where $e_i$ is the charge of the parton $i$ in electron charge units.
The compensation of the scale dependence takes place in
the following way. The $\mur$ scale dependence is compensated
in the expressions for the partonic cross sections: the scale
dependence of $\as$ in the Born term $\hat{\sigma}^{(0)}_{\gamma i}$
is compensated by the explicit scale dependence of the next-to-leading
term $\hat{\sigma}^{(1)}_{\gamma i}$. A similar cancellation occurs
in $\hat{\sigma}_{ij}$.
The dependence upon $\muf$ ($\muf^\prime$) cancels between the
explicit dependence in the next-to-leading order term and the
dependence in the structure functions convoluted with the Born
terms. This holds independently for the two terms of eq.~(\ref{sigmascales}).
The dependence upon $\mug$ cancels between the explicit dependence
in the next-to-leading order component of the first term of eq.
(\ref{sigmascales}) and the $\mug$ dependence of $f^{(\gamma)}_i$,
as given in eq. (\ref{inhomogeneous}), multiplied by the Born level
partonic cross section.
In the commonly-used photon density parametrizations,
$\mug$ is usually kept equal to $\muf$, so that the term given in
eq.~(\ref{inhomogeneous})
becomes a correction to the usual Altarelli-Parisi equation (the
so-called inhomogeneous term). Therefore, in our calculation, we
use for consistency $\muf^\prime=\mug$.
In some cases, the inclusion of the hadronic component gives only
a small effect, and will be neglected. In these cases we have chosen
$\mug=1\,$GeV, which amounts to setting the photon structure
function to zero at a scale of the order of a typical hadron mass.
We have found that varying $\mu_\gamma$ between 0.1 and 5~GeV does
not affect the results in a noticeable way.
\section{Total cross sections}
We begin our phenomenological study with the total cross sections for charm and
bottom production. We will concentrate here on the analysis of the dependence
of the total cross sections on the input parameters of the calculation, namely
the choice of parton distribution functions (PDFs), the choice of quark mass
and the choice  of the values of the scales discussed in Section~3. A
discussion of the ranges of masses and scales within which to explore the
dependence of the cross sections can be found in ref.~[\ref{MNR2}].

The target will always be an isosinglet nucleon, $N=(p+n)/2$, and unless
otherwise stated we will use the parton distribution set
MRSD0\refq{\ref{NEWMRS}}.

The default values of the charm and bottom mass will be 1.5 and 4.75 GeV
respectively, and the default choices for $\muf$ and $\mur$ will be:
\beq
\mur=m_c,\quad\muf=2m_c
\eeq
for charm and
\beq
\mur=\muf=m_b
\eeq
for bottom. The asymmetry in the default choice for the charm is related to the
scale threshold below which PDFs extrapolations are not available, as explained
in detail in ref.~[\ref{MNR2}].
As for $\mug$, we fix as a default $\mug=1$ GeV, as explained
in Section 3.

As an illustration of the reliability of the theoretical prediction
we present in fig.~\ref{ftot1} (fig.~\ref{ftot2}) the leading
and next-to-leading results for the total charm (bottom) cross
section.
The bands in figs.~\ref{ftot1} and \ref{ftot2} are obtained
by varying only the
renormalization and factorization scales, everything else being kept fixed.

Figures~\ref{ftot1} and \ref{ftot2} deserve some comments.
First of all, the scale uncertainty
associated with the charm production cross section is significantly smaller  at
the next-to-leading order for beam energies above 200 GeV, while below 200 GeV
the uncertainties at leading and next-to-leading order are similar. The
residual uncertainty is much smaller than in the case of hadroproduction for
comparable beam energies.
The contribution of the hadronic component of the photon is always smaller than
5\% of the total cross section for current energies.
Uncertainties coming from the determination of
the photon structure functions are therefore negligible with respect to others.

The reduction of the variation band is even more pronounced
in the case of bottom production,
once next-to-leading order corrections are included.
This is what we expect. For higher masses the value of $\as$ is smaller,
and the perturbative expansion becomes more reliable.
Observe that the size of the leading-order band
for the bottom cross section is not much smaller than the one for charm.
This is due to the fact that, as explained in ref.~[\ref{MNR2}], we did not
try to study the factorization scale dependence in the case of charm
production. The reader should therefore remember that a further uncertainty
should be added to the charm result, and that the band shown in
fig.~\ref{ftot1}
is only an underestimate of the uncertainties involved in the computation
of charm production cross sections.

We now turn to all other sources of uncertainties, such as the
structure function choice, the value of $\Lambda_{\sss QCD}$,
and the mass of the heavy
quarks.
In tables \ref{tcharm} and \ref{tbottom} we give the cross sections
for $\gamma$-nucleon collisions at various beam energies.
The rates were obtained using a reference scale $\mu_0 = m_c$ for charm
and $\mu_0=m_b$ for bottom.
We also show the effect
of varying $m_c$ between $1.2$ and $1.8$ GeV,
and $m_b$ between 4.5 and 5 GeV.
The scale $\mur$ was varied between
$\mu_0/2$ and $2\mu_0$. In the charm case, $\muf$ was kept equal to $2\mu_0$,
while for bottom we kept $\muf=\mur$. We verified that
considering independent variations for the factorization and renormalization
scale does not lead to a wider range in the bottom cross sections for the
energies shown in the tables.

The tables are broken into three blocks, each
corresponding to a different choice of $\Lambda_{\sss QCD}$
within the range allowed by
the current uncertainties. The upper block represents the default choice
relative to the MRSD0 fit. The second and third blocks correspond to the
sets discussed in ref.~[\ref{HMRS1}] for the nucleon.
The values of $\Lambda_4$
obtained in the fits of ref.~[\ref{HMRS1}] range
from 135 to 235 MeV,
corresponding to a range for $\Lambda_5$ between 84 and 155 MeV.
This range for $\Lambda_5$ is chosen because no good fit to deep
inelastic data is possible outside that range in the context
of ref.~[\ref{HMRS1}] (\ie\ with that choice of structure function
parametrization, etc.). We have chosen instead the wider
range\refq{\ref{AltarelliAachen}} $100<\Lambda_4<300\;$MeV,
corresponding to $60<\Lambda_5<204\;$MeV.
Therefore, in order to take into account the full range of uncertainty
associated with the value of $\Lambda_{\sss QCD}$, we were forced to
account only partially for the correlation beteween $\Lambda_{\sss QCD}$
and the nucleon structure functions.

For the charm cross section, as can be seen, the value of the charm quark mass
is the major source of uncertainty. Differences between the extreme choices
$\mc=1.2$ and 1.8 GeV vary from a factor of 5 at 100 GeV to a factor of 3 at
400
GeV. Differences due to the scale choice are of the order of a factor of 2 at
low energy, and at most 50\% at higher energy. A factor of 2 uncertainty also
comes from the variation of $\Lambda_{\sss QCD}$ within the chosen range.

We also explored independently the effect of varying $\mug$. Differences are
totally negligible for values 0.1 GeV $<\mug<$ 5 GeV, and are not included
in the tables.

All of these uncertainty factors are systematically a factor of 2 or more
smaller than in the case of hadroproduction. Notice however one pathology
encountered when combining the most extreme choices of mass ($\mc=1.2$ GeV),
scale ($\mur=\mc/2$) and $\Lambda_{\sss QCD}$ (MRS235, $\Lambda_5=204$ MeV):
the cross
section in this case decreases in the region 100 GeV $< E_\gamma<$
600 GeV.
This happens because for this particular choice of parameters the radiative
corrections become negative in part of this energy range.

Low-energy measurements of charm photoproduction cross sections
favour a mass value of
approximately 1.5 GeV. Previous comparisons with theory, however, were made
using a fixed value $\mur=2\mc$. We expect that once the uncertainty on $
\Lambda_{\sss QCD}$
will be reduced, reliance on the next-to-leading-order calculation
and the residual dependence on
$\mur$ should allow a determination of \mc\ to within 100--200 MeV.

The corresponding variations in the bottom case are smaller, in particular for
energies sufficiently above the production threshold. At $E_\gamma=100$ GeV,
where we observe the largest uncertainty in the perturbative calculation, we
should also expect large non-perturbative effects and therefore the
perturbative prediction is not fully reliable. Notice also from
fig.~\ref{ftot2}\ that the contribution of the hadronic component of the photon
represents, at low energy, a significant fraction of the total cross section.
The reason is that, close to the threshold, the photon-gluon fusion process is
suppressed by the small gluon density at large $x$, while production via the
hadronic component of the photon can proceed through a light valence-quark
annihilation channel.

For completeness, we also give the contribution of the hadronic component of
the photon in tables~\ref{tcharm_had} and \ref{tbottom_had}, evaluated using
the photon structure function set ACFGP-mc of ref.~[\ref{Aurenche}] and the set
LAC1 of ref.~[\ref{Abramowicz}]. As can be seen, at current fixed-target
energies, this contribution is small.

Our final prediction for the allowed range of charm and bottom production cross
sections, including the full variation due to the scale  choice, the value of
$\Lambda_{\sss QCD}$ and the nucleon structure functions is shown as a function
of the
beam energy in fig.~\ref{ftot3}, for $\mc=1.2, 1.5$ and 1.8 GeV, and for
$m_b=4.5,\,4.75,\,5\;$GeV. The small contribution of the hadronic
component is not included in the figure.

\begin{table}
\begin{tabular}{|l||c|c|c|c|c|c|c|c|c|c|} \hline
& \multicolumn{3}{c|}{$m_c=1.2$ GeV}
& \multicolumn{3}{c|}{$m_c=1.5$ GeV} & \multicolumn{3}{c|}{$m_c=1.8$ GeV}
\\ \hline
$E_b$ \ \ \ $\mu_{\rm R}=$ & $m_c/2$ &$m_c$ & 2$m_c$ &
  $m_c/2$ &$m_c$ & 2$m_c$ &
  $m_c/2$ &$m_c$ & 2$m_c$ \\ \hline \hline
\multicolumn{10}{|c|}{Nucleon PDF set MRSD0, $\Lambda_5=140$ MeV}
\\ \hline
 60 GeV & 1.442 & 0.937 & 0.715 & 0.533 & 0.357 & 0.275 & 0.203 & 0.138 & 0.107
  \\ \hline
100 GeV & 1.673 & 1.205 & 0.952 & 0.746 & 0.537 & 0.427 & 0.341 & 0.248 & 0.198
  \\ \hline
200 GeV & 1.826 & 1.506 & 1.238 & 0.979 & 0.776 & 0.640 & 0.525 & 0.414 & 0.342
  \\ \hline
400 GeV & 1.906 & 1.753 & 1.481 & 1.160 & 0.997 & 0.844 & 0.691 & 0.582 & 0.495
  \\ \hline\hline
\multicolumn{10}{|c|}{Nucleon PDF set MRS135, $\Lambda_5= 60$ MeV}
\\ \hline
 60 GeV & 0.753 & 0.627 & 0.535 & 0.312 & 0.255 & 0.216 & 0.130 & 0.105 & 0.089
  \\ \hline
100 GeV & 0.921 & 0.807 & 0.705 & 0.443 & 0.379 & 0.329 & 0.217 & 0.183 & 0.159
  \\ \hline
200 GeV & 1.093 & 1.016 & 0.910 & 0.602 & 0.543 & 0.483 & 0.338 & 0.300 & 0.266
  \\ \hline
400 GeV & 1.229 & 1.190 & 1.082 & 0.740 & 0.694 & 0.629 & 0.451 & 0.417 & 0.377
  \\ \hline\hline
\multicolumn{10}{|c|}{Nucleon PDF set MRS235, $\Lambda_5=204$ MeV} \\ \hline
 60 GeV & 2.794 & 1.280 & 0.886 & 0.874 & 0.467 & 0.332 & 0.304 & 0.176 & 0.127
  \\ \hline
100 GeV & 3.012 & 1.631 & 1.179 & 1.183 & 0.700 & 0.518 & 0.502 & 0.314 & 0.236
  \\ \hline
200 GeV & 2.923 & 2.015 & 1.530 & 1.474 & 1.002 & 0.776 & 0.751 & 0.522 & 0.409
  \\ \hline
400 GeV & 2.714 & 2.329 & 1.830 & 1.663 & 1.280 & 1.024 & 0.956 & 0.731 & 0.592
  \\ \hline\hline
\end{tabular}
\caption{\label{tcharm}
Total charm cross sections ($\mu$b) in $\gamma N$ collisions. point-like photon
contribution. Nucleon PDF set as indicated.}
\end{table}

\begin{table}
\begin{tabular}{|l||c|c|c|c|c|c|c|c|c|c|} \hline
& \multicolumn{3}{c|}{$m_b=4.5$ GeV}
& \multicolumn{3}{c|}{$m_b=4.75$ GeV} & \multicolumn{3}{c|}{$m_b=5$ GeV}
\\ \hline
$E_b$ \ \ \ $\mu=$ & $m_b/2$ &$m_b$ & 2$m_b$ &
  $m_b/2$ &$m_b$ & 2$m_b$ &
  $m_b/2$ &$m_b$ & 2$m_b$ \\ \hline \hline
\multicolumn{10}{|c|}{Nucleon PDF set MRSD0, $\Lambda_5=140$ MeV}
\\ \hline
100 GeV & 0.081 & 0.049 & 0.031 & 0.035 & 0.020 & 0.012 & 0.013 & 0.008 & 0.004
  \\ \hline
200 GeV & 1.092 & 0.832 & 0.615 & 0.702 & 0.525 & 0.383 & 0.445 & 0.327 & 0.235
  \\ \hline
400 GeV & 3.969 & 3.373 & 2.760 & 2.929 & 2.469 & 2.001 & 2.163 & 1.807 & 1.452
  \\ \hline\hline
\multicolumn{10}{|c|}{Nucleon PDF set MRS135, $\Lambda_5= 60$ MeV}
\\ \hline
100 GeV & 0.083 & 0.057 & 0.038 & 0.038 & 0.025 & 0.016 & 0.016 & 0.010 & 0.006
  \\ \hline
200 GeV & 0.962 & 0.785 & 0.618 & 0.634 & 0.509 & 0.395 & 0.413 & 0.326 & 0.250
  \\ \hline
400 GeV & 3.279 & 2.922 & 2.511 & 2.459 & 2.174 & 1.852 & 1.845 & 1.618 & 1.366
  \\ \hline\hline
\multicolumn{10}{|c|}{Nucleon PDF set MRS235, $\Lambda_5= 204$ MeV}
\\ \hline
100 GeV & 0.094 & 0.055 & 0.033 & 0.039 & 0.022 & 0.013 & 0.015 & 0.008 & 0.005
  \\ \hline
200 GeV & 1.290 & 0.951 & 0.683 & 0.823 & 0.596 & 0.422 & 0.518 & 0.368 & 0.257
  \\ \hline
400 GeV & 4.714 & 3.909 & 3.118 & 3.459 & 2.847 & 2.251 & 2.541 & 2.074 & 1.626
  \\ \hline\hline
\end{tabular}
\caption{\label{tbottom}
Total bottom cross sections (nb) in $\gamma N$ collisions. point-like photon
contribution. Nucleon PDF set as indicated.}
\end{table}

\begin{table}
\begin{tabular}{|l||c|c|c|c|c|c|c|c|c|c|} \hline
& \multicolumn{3}{c|}{$m_c=1.2$ GeV}
& \multicolumn{3}{c|}{$m_c=1.5$ GeV} & \multicolumn{3}{c|}{$m_c=1.8$ GeV}
\\ \hline
$E_b$ \ \ \ $\mu_{\rm R}=$ & $m_c/2$ &$m_c$ & 2$m_c$ &
  $m_c/2$ &$m_c$ & 2$m_c$ &
  $m_c/2$ &$m_c$ & 2$m_c$ \\ \hline \hline
\multicolumn{10}{|c|}{Photon PDF set ACFGP--mc}
\\ \hline
 60 GeV & 0.050 & 0.024 & 0.015 & 0.011 & 0.007 & 0.004 & 0.004 & 0.002 & 0.002
  \\ \hline
100 GeV & 0.096 & 0.045 & 0.027 & 0.022 & 0.013 & 0.008 & 0.007 & 0.005 & 0.003
  \\ \hline
200 GeV & 0.203 & 0.094 & 0.056 & 0.052 & 0.029 & 0.019 & 0.017 & 0.011 & 0.007
  \\ \hline
400 GeV & 0.376 & 0.175 & 0.105 & 0.109 & 0.059 & 0.038 & 0.039 & 0.024 & 0.016
  \\ \hline\hline
\multicolumn{10}{|c|}{Photon PDF set LAC1}
\\ \hline
 60 GeV & 0.040 & 0.022 & 0.014 & 0.010 & 0.006 & 0.004 & 0.004 & 0.002 & 0.002
  \\ \hline
100 GeV & 0.123 & 0.053 & 0.031 & 0.019 & 0.012 & 0.008 & 0.006 & 0.004 & 0.003
  \\ \hline
200 GeV & 0.543 & 0.205 & 0.112 & 0.075 & 0.038 & 0.023 & 0.017 & 0.011 & 0.007
  \\ \hline
400 GeV & 1.737 & 0.647 & 0.353 & 0.292 & 0.134 & 0.080 & 0.068 & 0.037 & 0.024
  \\ \hline\hline
\end{tabular}
\caption{\label{tcharm_had}
Contribution of the hadronic component of the photon to the total
charm cross sections ($\mu$b) in $\gamma N$ collisions.
Nucleon PDF set is MRSD0, photon PDF set as indicated.}
\end{table}

\begin{table}
\begin{tabular}{|l||c|c|c|c|c|c|c|c|c|c|} \hline
& \multicolumn{3}{c|}{$m_b=4.5$ GeV}
& \multicolumn{3}{c|}{$m_b=4.75$ GeV} & \multicolumn{3}{c|}{$m_b=5$ GeV}
\\ \hline
$E_b$ \ \ \ $\mu=$ & $m_b/2$ &$m_b$ & 2$m_b$ &
  $m_b/2$ &$m_b$ & 2$m_b$ &
  $m_b/2$ &$m_b$ & 2$m_b$ \\ \hline \hline
\multicolumn{10}{|c|}{Photon PDF set ACFGP--mc}
\\ \hline
100 GeV & 0.006 & 0.006 & 0.005 & 0.003 & 0.003 & 0.003 & 0.001 & 0.001 & 0.001
  \\ \hline
200 GeV & 0.051 & 0.059 & 0.056 & 0.035 & 0.040 & 0.038 & 0.024 & 0.027 & 0.026
  \\ \hline
400 GeV & 0.168 & 0.189 & 0.183 & 0.123 & 0.141 & 0.138 & 0.092 & 0.107 & 0.104
  \\ \hline\hline
\multicolumn{10}{|c|}{Photon PDF set LAC1}
\\ \hline
100 GeV & 0.005 & 0.005 & 0.004 & 0.002 & 0.002 & 0.002 & 0.001 & 0.001 & 0.001
  \\ \hline
200 GeV & 0.054 & 0.058 & 0.053 & 0.037 & 0.039 & 0.035 & 0.024 & 0.025 & 0.023
  \\ \hline
400 GeV & 0.163 & 0.195 & 0.187 & 0.124 & 0.146 & 0.140 & 0.095 & 0.110 & 0.105
  \\ \hline\hline
\end{tabular}
\caption{\label{tbottom_had}
Contribution of the hadronic component of the photon to the total
bottom cross sections (nb) in $\gamma N$ collisions.
Nucleon PDF set is MRSD0, photon PDF set as indicated.}
\end{table}
\section{Differential distributions}
In this section we present the results for the one- and two-particle
differential distributions (the latter will also be referred to as {\em
correlations} in the following) for charm and bottom production.
Experiments can never obtain a monochromatic photon beam, even though the
energy of the photon can often be measured event by event. In order to retain a
large statistics, therefore,
differential distributions are most often presented integrating over the full
photon beam spectrum, rather than at a fixed value of $E_\gamma$. While the
total cross section is a slowly varying function of $E_\gamma$, features of the
distributions can be affected by the convolution over the photon beam energy
spectrum. A meaningful study therefore requires a similar convolution to be
performed within the theoretical calculation.

We have chosen to perform our phenomenological study using the photon beam
energy spectra of the two experiments NA14/2 at CERN\refq{\ref{na14}} and E687
at FNAL\refq{\ref{e687}}. The spectra were kindly provided to us by the two
collaborations. Since the energy range of the two spectra are rather different,
this choice can give an idea of the energy dependence of our results. The
spectra are shown in fig.~\ref{fgammae} and can be parametrized as follows:
\beq
\frac{1}{N}\frac{dN}{dx}=
C\left[x^{a_1}(1-x)^{a_2}+Rx^{b_1}(1-x)^{b_2}\right],
\eeq
where $x=E_{\gamma}/E_{max}$, $E_{min}<E_{\gamma}<E_{max}$,
\beqn
&&E_{max}=440\;{\rm GeV},\;\; E_{min}=125\;{\rm GeV}
\nonumber \\
&&a_1=0.2294, \;\;\; a_2=2.9533
\nonumber \\
&&R=7.5019 \times 10^6,\;\;\; b_1=16.766, \;\;\; b_2=11.190
\eeqn
for E687, and
\beqn
&&E_{max}=400\;{\rm GeV},\;\; E_{min}=20\;{\rm GeV}
\nonumber \\
&&a_1=2.5987, \;\;\; a_2=4.7211
\nonumber \\
&&R=-1.1862,\;\;\; b_1=2.7133, \;\;\; b_2=4.9516
\eeqn
for NA14/2. The constant $C$ is chosen to normalize the distributions
to unity.
The results we present are obtained by convoluting the theoretical
distributions with the above beam shapes. We will refer to these, respectively,
as NA14 beam and E687 beam.

Needless to say, only a detailed simulation of the detector acceptances and
efficiencies can allow a complete comparison of our results with the
experimental findings. Therefore one should take the results presented here as
indicative of the most relevant features of the next-to-leading-order
calculation.
Additional non-perturbative effects such as fragmentation and intrinsic
momentum of the partons inside the hadrons will also be discussed at the end.

As in the case of the total cross sections, the distributions will be
calculated  using the parton distribution set MRSD0\refq{\ref{NEWMRS}} for the
nucleon, unless otherwise stated.  The default values of the masses will be
$m_c=1.5$ GeV and $m_b=4.75$ GeV. The default values of the factorization and
renormalization scales $\muf$ and $\mur$ will be:
\beq
\muf=2\mu_0,\quad \mur=\mu_0
\eeq
for charm, and
\beq
\muf=\mu_0,\quad \mur=\mu_0
\eeq
for bottom, where
\beq
\mu_0 = \sqrt{m^2+p_{\sss T}^2}
\eeq
for the one-particle distributions and
\beq
\mu_0 = \sqrt{m^2+\frac{p_{\sss T}^2+\bar{p}_{\sss T}^2}{2} }
\eeq
for correlations.
Here $p_{\sss T}$ and $\bar{p}_{\sss T}$ are the transverse momenta
of the heavy quark and antiquark, respectively.
In the case of correlations, a further ambiguity arises
in the choice of the scale
(described in detail in ref.~[\ref{MNR1}]), which has to do with the freedom
of choosing the same or different values of the scales
for the three-body event and for the
corresponding counter-events. The results presented here will always follow
the approach of recomputing the scale for the counter-events.

\subsection{Charm production}
We begin with one-particle inclusive rates.
On the left side of fig.~\ref{fincc}  we show the  inclusive \pt\
distribution of $c$ quarks in the case of the E687 and NA14 photon beams.
The solid lines represent the full next-to-leading-order result.
The dots give the leading-order
contribution rescaled by a constant factor.
A slight stiffening of the \pt\ distribution after radiative corrections is
observed.

On the right-hand side of fig.~\ref{fincc}  we show the inclusive \xf\
distribution for the full next-to-leading-order calculation, superimposed onto
the rescaled Born result. Our definition of \xf\ is
\beq
\xf=\frac{2 p^{\sss CM}_\parallel}{E_{\sss CM}},
\eeq
where CM refers to the centre of mass of the target and the tagged photon beam.
In other words, this centre of mass has a boost with respect to the
laboratory frame, which depends on the energy of the photon responsible for the
interaction;
$p^{\rm CM}_\parallel$ is the momentum projection on the beam
direction in the centre-of-mass frame, and $E_{\rm CM}$ is the
centre-of-mass total energy.
The fraction of $c$ quarks produced with positive \xf\ is larger
than 90\%, due to the hardness of the photon probe.
Notice however that next-to-leading-order corrections induce a softening of the
distribution, due to processes where a photon splits into a light-quark pair
and interacts with a light quark from the nucleon.
We verified that the inclusive \pt\ spectrum does not change in shape if we
restrict ourselves to quarks produced at positive \xf, which is the region
where data are usually collected by the experiments.

Next we consider $c\bar c$ correlations, starting from
the invariant mass of the heavy
quark pair, which is shown in fig.~\ref{fmqqc}.
As before, the continuous lines represent the
full result of the next-to-leading-order calculation,
while the dots are the rescaled Born
results. The lower lines are obtained by imposing $\xf>0$ for both quarks.
Notice that while at the inclusive level most of the $c$'s have positive \xf,
a large fraction of pairs with large invariant mass have either the $c$ or the
$\bar c$ at $\xf<0$, as one should expect.

The difference in rapidity between the quark and the antiquark and
the \xf\ of the pair are shown in figs.~\ref{fdyna14}\ and \ref{fdye687}.
A slight broadening at next-to-leading order is observed
for the rapidity correlation, while a
dramatic change is observed at next-to-leading order in the case of
the pair \xf\ distribution. This dramatic change can be traced
back to the particular kinematics of heavy-quark photoproduction.
In the Born approximation, the \xf\ of the heavy-quark pair is simply
related to its invariant mass
\beq
x_{\sss F}^{Q\bar{Q}}=1-\frac{M_{Q\bar{Q}}^2}{S}
\eeq
and is therefore peaked at $x_{\sss F}^{Q\bar{Q}}$ near 1.
At the next-to-leading level, it is the \xf\ of the $Q\bar{Q}g$ system
that will be peaked near 1. This means that the \xf\ of the heavy-quark pair
will be markedly degraded. The contribution of $\gamma
q$ fusion is even softer, because of the contribution of the photon
splitting into a light quark-antiquark pair.

We now consider correlations that are trivial at the leading order, namely the
difference in azimuth and total transverse momentum
of the quark pair, \ptg. At leading order these
distributions are delta functions centred respectively at $\Delta \phi=\pi$
and $\ptg=0$. Higher-order real corrections such as gluon radiation or
gluon splitting processes smear them. We plot these distributions
in fig.~\ref{fptgc}.
Even after the inclusion of higher-order effects, the azimuthal correlation
shows a strong peak at $\Delta \phi\approx\pi$.
Likewise, the \ptg\ distribution is
dominated by configurations with \ptg\ smaller than 1 GeV. In both cases, the
tails are higher at the higher energy of the E687 experiment.

For all the previous distributions, changing the values of the charm mass
and renormalization scale $\mur$ results in large differences in rates but
small
and easily predictable shape modifications. The pattern of these changes is
similar to what is observed in the case of fixed-target
hadroproduction\refq{\ref{MNR2}}.

\subsection{Bottom production}
The differential distributions for bottom production are shown in
figs.~\ref{fincb} to \ref{fdphib}.
The cross section at NA14 is very small. We nevertheless include their
distributions for completeness.

The $p_T^2$ (single-inclusive) and the \mqq\ distributions are broader than
the corresponding distributions for charm production. They are, however,
narrower than would be expected on the basis of simple scaling arguments.
This is because $b$ production at fixed-target energies is still too
close to the threshold, and thus constrained by phase-space effects
(this can also be noticed from the strong energy dependence of the shape
of the curves). For the same reason, the $\Delta y$ distribution is
narrower than in the charm case.

The \ptg\ and the $\Delta  \phi$ distributions are much narrower, as a
consequence of the smaller value of $\as$ at the $b$ mass, and of the
previously mentioned phase-space constraints. The \xf\ distribution is softer
for bottom than for charm, and in particular the fraction of $b$ quarks with
negative \xf\ is larger. This is because a harder parton from the nucleon is
required to reach the energy threshold for the creation of a $b$ pair.

\subsection{Higher-Order Corrections and Hadronization}
The results discussed so far were obtained with a purely perturbative
calculation limited to the next-to-leading order. In the case of charm quarks,
the dependence of the cross section on the renormalization and factorization
scales indicates that higher-order corrections might be large. Nevertheless,
the stability of the shapes under inclusion of the next-to-leading-order terms
suggests that no significant changes should be expected in the differential
distributions when yet higher-order terms are included. This is not necessarily
true of possible non-perturbative corrections, such as the intrinsic \pt\ of
the initial-state partons, hadronization and fragmentation.  In particular, in
the regions of phase space close to threshold, or for $\ptg\to 0$, we should
expect significant corrections.

In our previous study of heavy-quark correlations in
hadroproduction\refq{\ref{MNR2}} we explored these effects using different
phenomenological models. In particular, we considered the parton shower Monte
Carlo HERWIG\refq{\ref{herwig}} to simulate both the backward evolution of the
initial state and the formation of charmed hadrons. We found that the backward
evolution of initial-state gluons gave rise to an artificially large intrinsic
\pt\ of the gluons themselves ($\langle k_{\rm{T}} \rangle$ = 1.7 GeV). This
large intrinsic \pt\ would significantly broaden the \dphi\ correlation, in
addition to softening the inclusive \pt\ distribution. We also found that, as
expected, the colour singlet cluster formation and decay into hadrons leads to
a large colour drag in the direction of the hadron beam. The combination of the
intrinsic \pt\ smearing, of the colour drag, and of the decay of unstable
charmed hadrons, resulted in inclusive \pt\ and \xf\ distributions of $D$
mesons that are slightly harder than those resulting from the purely
perturbative calculation. The \dphi\ correlations remain significantly broader
than described by next-to-leading-order QCD. Both these results are supported
by current experimental evidence\refq{\ref{appel}}.

In ref.~[\ref{MNR2}] we were also able to parametrize the effect of
the intrinsic
\pt\ by using a Gaussian smearing, while we argued that there is no solid basis
for the application of the Peterson formalism\refq{\ref{peterson}}
to describe the fragmentation in the large-\xf\ region. In this section, we
repeat the analysis in the case of heavy-quark photoproduction, for
the case of charm production with the E687 photon spectrum.

Figure~\ref{fherpt} shows a comparison of inclusive \pt\ and \xf\ distributions
obtained from the perturbative calculation (solid lines), from HERWIG before
hadronization (dotted lines), and from HERWIG after hadronization (dashed
lines). In the case of the inclusive \pt\ distribution we find that, as
observed in hadroproduction, the effects of intrinsic \pt\ and of hadronization
tend to respectively harden and soften the distribution. The net
result is a softening of the shape.
This is due to the fact that only one initial-state parton (namely the gluon)
can acquire a transverse \pt\ in photoproduction, contrary to hadroproduction
where we have two gluons in the initial state and the
effect is enhanced.

The most dramatic effects are however observed in the inclusive \xf\
distribution. Hadronization effects heavily suppress the
production of charmed hadrons at large \xf, which was favoured at
the purely perturbative level. This is perhaps surprising, because it would
be expected that for heavy quarks produced in the photon fragmentation region
we should be able to obtain the correct distribution by using the
perturbative calculation, convoluted with the fragmentation
function for the $c$-quark fragmenting into a $D$ meson.
This is not necessarily true. As
in the hadroproduction case, there are no theoretical reasons to
support this possibility. The fragmentation function formalism is
in fact applicable only when the elementary production
process takes place at energies much larger than the mass of the heavy
quark. This is the case, for example, in the production of heavy
flavoured mesons in $e^+e^-$ collisions, or in hadroproduction and
photoproduction at high transverse momentum. The factorization theorem
states that the same fragmentation function, evolved to the
appropriate scale, and convoluted with the
perturbative calculation of the partonic subprocess, should describe
all these processes.
The \xf\ spectrum, instead, is not really characterized by a
high-energy elementary process. The heavy-quark pair is produced with
a relatively small invariant mass, and the large \xf\ region is
reached when the production angle in the heavy-quark centre of mass
is small. Under these circumstances, non-perturbative effects (other
than the fragmentation effects) could also take place. For example,
the heavy quark could feel the dragging of the heavy antiquark, or of
the beam remnants.

Figure \ref{fherptg} shows the HERWIG results for the charm pair \ptg\ and
\dphi\ distributions. While the \dphi\ correlation is still significantly
broader than that calculated at the perturbative level, the distribution is
more peaked than the one evaluated from HERWIG in the hadroproduction case.
This results from the smaller effect of intrinsic \pt\ present in the
photoproduction, as already mentioned above. Existing data
\refq{\ref{na14},\ref{e687}}\ agree qualitatively with this result.

In fig.~\ref{fptk}, finally, we show the charm pair \ptg\ and
\dphi\ distributions obtained by
giving a random intrinsic transverse momentum to the incoming gluon,
with a Gaussian distribution\refq{\ref{MNR2}},
for different values of $\langle p^2_{\sss T } \rangle$.
As can be seen, the choice
$\langle k_{\sss T }^2 \rangle = 3$~GeV$^2$ reproduces quite closely the HERWIG
result for the \dphi\ correlation. As already discussed in ref.~[\ref{MNR2}],
we
expect such a large intrinsic \pt\ to be a pure artefact of the Monte Carlo. If
the data were to confirm the existence of broad \dphi\ correlations as shown in
fig.~\ref{fherptg},  it might be interesting to try to justify theoretically on
a more solid basis the possibile existence of such a large intrinsic \pt\ for
the gluons inside the hadron.

\subsection{Effect of the photon hadronic component}
As already seen in Section~4, the effect of the hadronic
component on the total cross section,
predicted using standard photon structure functions, is generally
small. The question now arises whether its
effect on the shape of distributions is also negligible.
We have studied this problem using the next-to-leading calculation
of ref.~[\ref{MNR1}], together with the parametrization of the photon
structure function set ACFGP-mc of ref.~[\ref{Aurenche}], and the set LAC1 of
ref.~[\ref{Abramowicz}], at a fixed photon energy of 230~GeV.
We have found that the only important
modifications occur in the \xf\ distribution for a single heavy quark,
and for the \xf\ of the pair. In the inclusive \xf\ distribution for charm,
we find that the hadronic component becomes comparable to the
point-like term for $x_{\sss F}\approx -0.3$, and for smaller \xf\ it remains
of the same order. For $b$ production, we find instead that the
hadronic component becomes comparable to the point-like term
for $x_{\sss F} \approx -0.7$, a region in which the cross section
is several orders of magnitude below the peak value.

For the \xf\ of the pair, the effect of the hadronic component is
more pronounced. This is due to the fact that the point-like component
is concentrated near $x^{\sss Q\bar{Q}}_{\sss F}=1$. On the other hand
the hadronic component is distributed in the central region
(we find that for charm it peaks at $x_{\sss F}\approx -0.15$ with
the LAC1 set, and at $x_{\sss F}\approx 0$ with the ACFGP-mc set).
Its contribution in this region is therefore of the magnitude given
in tables~\ref{tcharm_had} and \ref{tbottom_had}. Observe, however,
that (as discussed in the previous subsection) hadronization effects
do spread out the \xf\ distribution of the pair. It is unlikely,
therefore, that one can use these distributions to make statements
about the hadronic component of the photon.

We conclude, therefore, that for all practical purposes, the
hadronic component can be neglected altogether in the fixed-target
experimental configurations of present interest.
\section{Conclusions}
We have performed a calculation of next-to-leading-order QCD corrections
to heavy-quark production in photon-hadron collisions.
Our calculation improves over previous results, in that it
can be used to compute
any distribution in the heavy quark and antiquark, and possibly
in the extra jet variables.

 We have presented
a phenomenological study of total cross sections, single-inclusive
distributions and correlations, for charm and bottom production
at fixed-target energies.

A detailed study of all the theoretical uncertainties of the calculation,
namely those due to an independent variation of heavy-quark mass,
factorization and renormalization scales and $\Lambda_{\sss QCD}$,
has been performed for the total cross sections. We found that
the next-to-leading contribution is less important here than in the
hadroproduction case. Therefore, we expect the full result to be more
reliable for photoproduction than for hadroproduction of heavy flavours.

For single and double differential distributions, in the case of
charm production, we always find that non-perturbative effects could
be important. For the \pt\ distribution, a possible description
of the non-perturbative effects is given via the introduction
of an intrinsic transverse momentum for the incoming gluon, which
tends to stiffen the \pt\ distribution, together with a fragmentation
function similar to the ones used in $e^+e^-$ physics. This seems to
give a reasonable description of the \pt\ and \dphi\ distributions
in both photoproduction and hadroproduction.

In the case of the \xf\ distribution, it is very difficult to
describe the non-perturbative effects in a simple way. From
Monte Carlo studies, we conclude that it is likely
that colour-dragging effects prevail, thus making it difficult to give
a homogeneous description of the hadroproduction and photoproduction
\xf\ distribution.
\newpage
\appendix
\section*{Soft limit of the real amplitude}
In this appendix we present the calculation of the
amplitude for the process $\gamma g\to Q\bar{Q} g$ in the limit when
the momentum of the emitted gluon tends to zero. This is the
only case of interest, because the analogous limit for
the process $\gamma q\to Q\bar{Q} q$ gives a trivial result.
Momenta and colour and polarization indices are assigned
as follows:
\beq
\label{momenta}
\gamma (p_1,\mu)\,+\, g(p_2,\nu,a)\,\rightarrow\,
Q(k_1,i)\,+\,\bar{Q}(k_2,j)\,+\, g(k,\rho,b)\, .
\eeq
The main difference with respect to the heavy-quark hadroproduction
case is that the gluon cannot directly couple to the incoming photon.
In principle we are then left with three potentially singular diagrams
when $k\to 0$, namely the diagrams in which the outgoing gluon is emitted
by the incoming parton or by the heavy quark or antiquark.
We indicate the amplitude
for the process $\gamma g\to Q\bar{Q}$ with
\beq
\bar{u}(k_1)\, {\rm A}^a_{\mu\nu ;ij}\, v(k_2)\, ,
\eeq
where the momenta and indices are as in eq.~(\ref{momenta}).
It then follows that the contribution
of the diagram in which the gluon is emitted
from the outgoing heavy quark is given by
\beq
\label{diag1}
\frac{g_{\sss S}\lambda^b_{il}}{2}\, \bar{u}(k_1)\,\gamma_\rho\,
\frac{\hat{k}+\hat{k}_1+m}{(k+k_1)^2 - m^2}\,
{\rm A}^a_{\mu\nu ;lj}\, v(k_2)\, .
\eeq
When the gluon is emitted from the antiquark we have instead
\beq
\label{diag2}
\frac{g_{\sss S}\lambda^b_{lj}}{2}\, \bar{u}(k_1)\,
{\rm A}^a_{\mu\nu ;il}\, \frac{\hat{k}+\hat{k}_2+m}{(k+k_2)^2 - m^2}\,
\gamma_\rho\, v(k_2)\, .
\eeq
Finally, when the gluon is emitted from the incoming gluon leg the
contribution is
\beq
\label{diag3}
i g_{\sss S}\,\frac{g^{\sigma\lambda}}{(p_2-k)^2}\,
G^{abc}_{\nu\rho\sigma}(p_2,-k,k-p_2)\,
\bar{u}(k_1)\, {\rm A}^a_{\mu\lambda ;ij}\, v(k_2)\, ,
\eeq
where
\beq
G^{abc}_{\mu_1\mu_2\mu_3}(q_1,q_2,q_3)\,=\,f^{abc}\left[
g_{\mu_1\mu_2}(q_1-q_2)_{\mu_3} +
g_{\mu_2\mu_3}(q_2-q_3)_{\mu_1} +
g_{\mu_3\mu_1}(q_3-q_1)_{\mu_2} \right]
\eeq
is the quantity that appears in the QCD three-gluon vertex.
We can now evaluate the $k\to 0$ limit in eqs.~(\ref{diag1}),
{}~(\ref{diag2}) and ~(\ref{diag3}). The soft
limit of the three-body real amplitude is
\beqn
\label{softamp}
{\cal A}^{(r)}_{\gamma g}\left(k\to 0\right)\,&=&\,
\frac{g_{\sss S}\lambda^b_{il}}{2} \frac{k_{1\rho}}{k\cdot k_1}\,
\bar{u}(k_1)\, {\rm A}^a_{\mu\nu ;lj}\, v(k_2)
-\frac{g_{\sss S}\lambda^b_{lj}}{2} \frac{k_{2\rho}}{k\cdot k_2}\,
\bar{u}(k_1)\, {\rm A}^a_{\mu\nu ;il}\, v(k_2)
\nonumber \\
&&+ i g_{\sss S}f^{abc} \frac{p_{2\rho}}{k\cdot p_2}\,
\bar{u}(k_1)\, {\rm A}^c_{\mu\nu ;ij}\, v(k_2)\, .
\nonumber \\
\eeqn
At the lowest order, the amplitude for the two-body process
$\gamma g\to Q\bar{Q}$ can be written as
\beq
\bar{u}(k_1)\, {\rm A}^a_{\mu\nu ;ij}\, v(k_2)\equiv
\lambda^a_{ij} {\rm B}_{\mu\nu}\,.
\eeq
Using the identity
\beq
2if^{abc}\lambda^c_{ij}\,=\,\left(\lambda^a\lambda^b\right)_{ij}
- \left(\lambda^b\lambda^a\right)_{ij}
\eeq
eq.~(\ref{softamp}) becomes
\beq
{\cal A}^{(r)}_{\gamma g}\left(k\to 0\right)\,=\,
\frac{g_{\sss S}}{2}\left[\left(\lambda^b\lambda^a\right)_{ij}
\left(\frac{k_{1\rho}}{k\cdot k_1} - \frac{p_{2\rho}}{k\cdot p_2}\right)
-\left(\lambda^a\lambda^b\right)_{ij}
\left(\frac{k_{2\rho}}{k\cdot k_2} - \frac{p_{2\rho}}{k\cdot p_2}\right)
\right] {\rm B}_{\mu\nu}\, .
\eeq
Squaring and summing over initial and final degrees of freedom
we obtain
\beqn
\abs{{\cal A}^{(r)}_{\gamma g}\left(k\to 0\right)}^2&=&
\frac{g^2_{\sss S}}{4}\Bigg[
{\rm Tr}\left(\lambda^b\lambda^a\lambda^a\lambda^b\right)
\Bigg((k_1 k_1)-2(p_2 k_1)-2(p_2 k_2)+(k_2 k_2)\Bigg)
\nonumber \\&&
-2{\rm Tr}\left(\lambda^b\lambda^a\lambda^b\lambda^a\right)
\Bigg((k_1 k_2)-(p_2 k_1)-(p_2 k_2)\Bigg)\Bigg]
\sum_{spin}{\rm B}_{\mu\nu}\left({\rm B}^{\mu\nu}\right)^\dagger \, ,
\nonumber \\
\eeqn
where we introduced the eikonal factors defined by
\beq
(v w)\,=\,\frac{v \cdot w}{v\cdot k\,\, w\cdot k}\, ,
\eeq
and we used the masslessness of the incoming parton, $p_2^2=0$.
Evaluating the traces
\beq
{\rm Tr}\left(\lambda^b\lambda^a\lambda^a\lambda^b\right)\,=\,
16 T_{\sss F} D_{\sss A} C_{\sss F}\, ,\;\;\;\;\;\;
{\rm Tr}\left(\lambda^b\lambda^a\lambda^b\lambda^a\right)\,=\,
16 T_{\sss F} D_{\sss A} \left(C_{\sss F}-\frac{1}{2}C_{\sss A}\right)\, ,
\eeq
inserting the proper flux factor, and averaging, we finally have
the squared three-body amplitude in the soft limit
\beqn
\label{finale}
{\cal M}^{(s)}_{\gamma g}\,&=&\,
g^2_{\sss S}\Bigg[C_{\sss F}\Bigg((k_1 k_1)+(k_2 k_2)\Bigg)
-2\left(C_{\sss F}-\frac{1}{2}C_{\sss A}\right)(k_1 k_2)
\nonumber \\&&
-C_{\sss A}\Bigg((p_2 k_1)+(p_2 k_2)\Bigg)\Bigg]
{\cal M}^{(b)}_{\gamma g}\, .
\eeqn
For QCD with three colours $T_{\sss F}=1/2$, $D_{\sss A}=8$,
$C_{\sss F}=4/3$ and $C_{\sss A}=3$. Here
\beq
{\cal M}^{(b)}_{\gamma g}\,=\,\frac{1}{2s}\frac{1}{4 D_{\sss A}}
{\rm Tr}\left(\lambda^a\lambda^a\right)
\sum_{spin}{\rm B}_{\mu\nu}\left({\rm B}^{\mu\nu}\right)^\dagger
\eeq
is the Born amplitude squared for the two-body process $\gamma g\to Q\bar{Q}$.
By direct calculation we obtain
\beq
{\cal M}^{(b)}_{\gamma g}\,=\,
\frac{e^2_{\sss Q} g^2_{\sss S} T_{\sss F}}{sut}
\left[\left(t^2+u^2-\ep s\right)\left(1-\ep\right)
+4 m^2 s\left(1-\frac{m^2 s}{ut}\right)\right]\, ,
\eeq
where $e_{\sss Q}$ is the heavy-quark charge.
Equation (\ref{finale}) can now be integrated over three-body phase space.
Using the notations of Section 2 we have
\beqn
\fpgs\left(\thu\right)&=&\int dx\, dy\, d\thd\, \delta (1-x)
\left(1-y^2\right)^{-1-\ep}\sin^{-2\ep}\thd
\left[4 t_k u_k \mpg\right]
\nonumber \\
&=&\int dx\, dy\, d\thd\, \delta (1-x)
\left(1-y^2\right)^{-1-\ep}\sin^{-2\ep}\thd
\left[4 t_k u_k \mpg^{(s)}\right]\, .
\eeqn
The only non-trivial point of the calculation is the integration of the eikonal
factors
\beq
I_{\left(v w\right)}=\int dy\, d\thd\,
\left(1-y^2\right)^{-1-\ep}\sin^{-2\ep}\thd
\left[4 t_k u_k \left(v w\right)\right]_{x=1}\, .
\eeq
The expression for $\fpgs\left(\thu\right)$ is obtained
by formal substitution of each $\left(v w\right)$ eikonal factor
in eq.~(\ref{finale}) for the corresponding integral
$I_{\left(v w\right)}$. We give here all the eikonal integrals,
thus correcting some misprints of the analogous formulae in
ref.~[\ref{MNR1}].
\beqn
I_{(p_1 p_2)} &=& -8s\frac{\pi}{\ep}
\nonumber\\
I_{(p_1 k_1)} &=&  4s\pi
\Bigg[-\frac{1}{\ep}+\vltm+\ltuno-\frac{\ep}{2}\vlpmq
\nonumber \\
&&-2\ep \, \litwo\left(1+\frac{2t}{s(1-\beta)}\right)
   -2\ep \, \litwo\left(1+\frac{2t}{s(1+\beta)}\right)
\nonumber \\
&&
-2\ep\log\frac{-2t}{s(1+\beta)}\log\frac{-2t}{s(1-\beta)} \Bigg]
\nonumber\\
I_{(p_2 k_1)} &=& I_{(p_1 k_1)} (t \to u)
\nonumber\\
I_{(p_1 k_2)}&=&I_{(p_2 k_1)}
\nonumber\\
I_{(p_2 k_2)}&=&I_{(p_1 k_1)}
\nonumber\\
I_{(k_1 k_1)}&=&8 s \pi \left(1+\frac{\ep}{\beta}\,\vlpm\right)
\nonumber\\
I_{(k_2 k_2)}&=&I_{(k_1 k_1)}
\nonumber\\
I_{(k_1 k_2)}&=&8 s \left(1-\frac{\rho}{2}\right)
                \frac{\pi}{\beta} \left[\vlpm+\ep\left(\softb\right)\right].
\eeqn
\noindent{\bf Acknowledgements}

\noindent We are very grateful to the members of the NA14/2 and
E687 collaborations for providing us with relevant material. We are
thankful to J.~Butler, R.~Gardner, J.~Wiss and particularly to J.~Appel
for several discussions and for twisting our arm to carry this project to
completion.

\begin{reflist}
\item\label{HeraWorkshop}
   For a review and a complete bibliography, see ``Physics at HERA'',
   Proceedings of the Workshop,
   DESY, Hamburg, eds. W. Buchm\"uller and G. Ingelman (1991).
\item\label{EllisNason}
   R.K.~Ellis and P.~Nason, \np{B312}{89}{551}.
\item\label{SmithNeerven}
   J.~Smith and W.L.~van Neerven, \np{B374}{92}{36}.
\item\label{expt}
   M.I.~Adamovich et al. (Photon Emulsion Collaboration), \pl{B187}{87}{437}
   and references therein;\\
   J.C.~Anjos et al. (E691 Collaboration), \prl{65}{90}{2503};\\
   R.W.~Forty (NA14$^\prime$ Collaboration), in the Proceedings of the
   XXIV International Conference on High Energy Physics, Munich,
   eds. R.~Kotthaus and J.H.~K\"uhn (1988);\\
   D.~Buchholz et al. (E687 Collaboration), in the Proceedings of the
   XXV International Conference on High Energy Physics, Singapore,
   eds. K.K. ~Phua and Y.~Yamaguchi (1990).
\item\label{MNR1}
   M.L.~Mangano, P.~Nason and G.~Ridolfi, \np{B373}{92}{295}.
\item\label{MNR2}
   M.L.~Mangano, P.~Nason and G.~Ridolfi, IFUP-TH-37-92, to appear
   in Nucl. Phys. B.
\item\label{ZZ}
   B.~Mele, P.~Nason and G.~Ridolfi, \np{B357}{91}{409}.
\item\label{FMNR1}
   S.~Frixione, M.L.~Mangano, P.~Nason and G.~Ridolfi,
   CERN-TH.6864/93 (1993), to appear in Phys. Lett. B.
\item\label{FMNR3}
   S.~Frixione, M.L.~Mangano, P.~Nason and G.~Ridolfi,
   in preparation.
\item\label{Factorization}
See, e.g., J.C. Collins, D.E. Soper and G. Sterman, in {\it Perturbative QCD}
	(Singapore: World Scientific, 1989), p. 1.
\item\label{BBDM}
   W.~A.~Bardeen, A.~Buras, D.~W.~Duke and T.~Muta, \pr{D18}{78}{3998}.
\item\label{EllisKunszt}
   R.K.~Ellis and Z.~Kunszt, \np{B303}{88}{653}.
\item\label{NEWMRS}
   A.D. Martin, R.G. Roberts and W.J. Stirling,
   \pr{D47}{93}{867}.
\item \label{Aurenche}
   % funzioni di struttura del fotone
   P. Aurenche, P. Chiappetta, M. Fontannaz, J.P. Guillet and E. Pilon,
   \zp{C56}{92}{589}.
\item\label{Abramowicz}
   H. Abramowicz, K. Charchula and A. Levy, \pl{269B}{91}{458}.
\item \label{HMRS1}
	A.D. Martin, R.G. Roberts and W.J.~Stirling,
	\pr{D43}{91}{3648}.
\item \label{AltarelliAachen}
        G. Altarelli, CERN-TH. 6623/92.
\item\label{na14}
   M.P.~Alvarez et al. (NA14/2 Collaboration), preprint CERN-PPE/92-28
   (1992) and \pl{B278}{92}{385}.
\item\label{e687}
   P.L.~Frabetti et al. (E687 Collaboration), preprint
   FERMILAB-PUB-93-072-E (1993).
\item \label{herwig}
	G. Marchesini and B.R. Webber, \np{B310}{88}{461}.
\item  \label{appel}
	J. Appel, {\em Annu. Rev. Nucl. Part. Sci.}, vol. 42 (1992).
\item \label{peterson}
	C. Peterson, D. Schlatter, I. Schmitt and P. Zerwas, \pr{D27}{83}{105}.
\end{reflist}
\begin{figcap}
%-----------------------------------------------------------------------
% FPHCTOT.TOP
	\item  	\label{ftot1}
Total cross section at leading and next-to-leading order
for charm production in $\gamma N$ collisions as a
function of the beam energy. We plot the range of variation for the scale
changes indicated in the figure, for $m_c=1.5$ GeV.
MRSD0 parton distribution set. The dotted line represents the contribution from
the hadronic component of the photon, evaluated using
the set ACFGP-mc.
%-----------------------------------------------------------------------
% FPHBTOT.TOP
	\item  	\label{ftot2}
Total cross section at leading and next-to-leading order
for bottom production in $\gamma N$ collisions as a
function of the beam energy. We plot the range of variation for the scale
changes indicated in the figure, for $m_b=4.75$ GeV.
MRSD0 parton distribution set.
%-----------------------------------------------------------------------
% FPHTOTSC.TOP
	\item  	\label{ftot3}
Total cross section for charm and bottom production in $\gamma N$ collisions as
a function of the beam energy. We plot the overall range of variation for
changes in the parameters as discussed in the text, each band representing the
result for a specified value of the quark mass.
%-----------------------------------------------------------------------
% NON ESISTE ANCORA !!!
	\item	\label{fgammae}
Photon beam energy spectra for NA14/2 and for E687.
%-----------------------------------------------------------------------
% FPHPTXFC.TOP
	\item  	\label{fincc}
Charm inclusive $\pt$ and $\xf$ distributions in $\gamma N$ collisions with the
E687 and NA14 photon beam energy spectrum.
%-----------------------------------------------------------------------
% FPHMQQC.TOP
	\item  	\label{fmqqc}
Invariant-mass distribution of charm pairs produced in $\gamma
N$ collisions with the E687 and NA14 photon beam energy spectrum.
%-----------------------------------------------------------------------
% FPHDYXFQQCNA14.TOP
	\item  	\label{fdyna14}
Rapidity correlation and $\xf$ distribution for charm pairs produced in $\gamma
N$ collisions with the NA14 photon beam energy spectrum. The lower curves are
obtained requiring both quarks to have $\xf>0$.
%-----------------------------------------------------------------------
% FPHDYXFQQCE687.TOP
     	\item  	\label{fdye687}
Same as fig.~\ref{fdyna14}, but for the E687 photon beam spectrum.
%-----------------------------------------------------------------------
% FPHPTQQPHIC.TOP
	\item  	\label{fptgc}
Charm pair \ptg\ and azimuthal correlation in $\gamma N$ collisions with the
E687 and NA14 photon beam spectrum.
%-----------------------------------------------------------------------
% FPHPTXFB.TOP
	\item  	\label{fincb}
Bottom inclusive $\pt$ and $\xf$ distributions in $\gamma N$ collisions with
the
E687 and NA14 photon beam spectrum.
%-----------------------------------------------------------------------
% FPHPTQQMQQB.TOP
	\item  	\label{fmqqb}
\ptg\ and invariant-mass distributions of bottom pairs produced in $\gamma
N$ collisions with the E687 and NA14 photon beam spectrum.
%-----------------------------------------------------------------------
% FPH3B.TOP
	\item  	\label{fdphib}
Bottom pair correlations in $\gamma N$ collisions with the E687 (solid lines)
and NA14 (dashed lines) photon beam energy spectrum: azimuthal correlations
(left side), \xf\ of the pair (right side) and $b\bar b$ radipity difference
(left inset).
%-----------------------------------------------------------------------
% FPHINCHRW.TOP
  	\item	\label{fherpt}
Comparison between HERWIG and the ${\cal O}(\aem\as^2)$ result
(solid) for inclusive distributions of charm with the E687 photon beam.
For HERWIG we plot the variables relative to the charm quark
before hadronization (dotted line) and relative to stable charm hadrons
(dashed line).
%-----------------------------------------------------------------------
% FPHPTHRW.TOP
	\item 	\label{fherptg}
Comparison between HERWIG and the ${\cal O}(\aem\as^2)$ result (solid) for
inclusive \ptg\ and \dphi\ distributions of charm with the E687 photon beam.
Different line patterns are explained in the previous figure's caption.
%-----------------------------------------------------------------------
% FPHPTK.TOP
\item 	\label{fptk}
Effect of a non-perturbative $p_T$ kick for the incoming parton in the nucleon,
compared with the ${\cal O}(\aem\as^2)$ effect. The curves with a $p_T$ kick
are
obtained with the Born cross section with MRSD0 structure functions,
supplemented by a random $p_T$ kick on the incoming
parton. The NLO curves are obtained with the same structure functions  and
were rescaled to the same  normalization as the other curves.
\end{figcap}
\end{document}